# PLANT AND CONTROLLER OPTIMIZATION FOR POWER AND ENERGY SYSTEMS WITH MODEL PREDICTIVE CONTROL

**Donald J. Docimo**[*]
Department of Mechanical Engineering, Texas Tech University
Lubbock, TX, U.S.A.
donald.docimo@ttu.edu

**Ziliang Kang**
Department of Aerospace Engineering, University of Illinois at Urbana-Champaign
Urbana, IL, U.S.A.
kang134@illinois.edu

**Kai A. James**
Department of Aerospace Engineering, University of Illinois at Urbana-Champaign
Urbana, IL, U.S.A.
kaijames@illinois.edu

**Andrew G. Alleyne**
Department of Mechanical Science and Engineering, University of Illinois at Urbana-Champaign
Urbana, IL, U.S.A.
alleyne@illinois.edu

## ABSTRACT

*This article explores the optimization of plant characteristics and controller parameters for electrified mobility. Electrification of mobile transportation systems, such as automobiles and aircraft, presents the ability to improve key performance metrics such as efficiency and cost. However, the strong bidirectional coupling between electrical and thermal dynamics within new components creates integration challenges, increasing component degradation and reducing performance. Diminishing these issues requires novel plant designs and control strategies. The electrified mobility literature provides prior studies on plant and controller optimization, known as control co-design (CCD). A void within these studies is the lack of model predictive control (MPC), recognized to manage multi-domain dynamics for electrified systems, within CCD frameworks. This article addresses this through three contributions. First, a thermo-electro-mechanical hybrid electric vehicle (HEV) model is developed that is suitable for both plant optimization and MPC. Second, simultaneous plant and controller optimization is performed for this multi-domain system. Third, MPC is integrated within a CCD framework using the candidate HEV model. Results indicate that optimizing both the plant and MPC parameters simultaneously can reduce physical component sizes by over 60% and key performance metric errors by over 50%.*

## I. INTRODUCTION

This article addresses plant and controller optimization for power and energy systems. There is an ongoing shift in the expectations of mobile power systems such as vehicles and aircraft, which has promoted the development of partially or fully electrified vehicle architectures. In the U.S. transportation sector, electricity as an energy source is predicted to grow, on average, 7.4% per year up to 2050 [1]. Electrification of these systems is motived by its potential to meet new metrics: (a) Reduce greenhouse gas emissions [2] to meet future standards set by different countries [3]; (b) increase efficiency [4,5]; and (c) minimize fuel cost [5]. The electrification of mobility and transportation systems necessitates the introduction of new components into conventional designs. The dynamics of these new electrical components are strongly coupled to temperature [6–8], triggering integration issues that complicate overall performance and component degradation [9,10].

Offsetting these integration issues requires development of new (i) real-time control algorithms and (ii) plant design methodologies to maximize closed-loop plant performance for these dynamically complex systems. An additional possibility is the ability to develop both of these simultaneously, known as (iii) control co-design. As presented below, prior literature has explored several of these areas, usually in isolation from the overall system evaluation.

*Advancements in real-time control.* The automotive literature presents several alternative strategies for energy management of powertrains with electro-mechanical dynamics. This includes iterative dynamic programming [11], stochastic dynamic programming (SDP) [12,13], equivalent consumption minimization strategies [14,15], and rule-based strategies [16,17], as well as strategies enhanced by machine learning [18] and data-driven methods [19]. Some of these strategies have been extended to include consideration of component health [20], such as using SDP to match desired outputs while minimizing energy cost and energy storage device degradation

---

[*] Address all correspondence to this author.

[21]. Building on this, recent studies include thermal dynamics of the cabin [22] and battery pack [23] under the supervision of the energy management controller. This allows for the consideration of the bidirectional coupling between electrical and thermal dynamics in hybrid electric vehicles (HEVs) and electric vehicles (EVs), mitigating integration issues. Model-based algorithms such as model predictive control (MPC) are an alternative option to manage multi-domain dynamics of vehicles [24,25]. This includes hierarchical and distributed MPC architectures, which are particularly capable of managing timescale separation caused by fast electrical and slow thermal dynamics [26–29].

*Advancements in plant design.* Plant design can be categorized by studies focusing on sizing optimization and topology optimization. Sizing optimization seeks to optimize component sizes within the system, including thermal component design [30] and electro-mechanical components [31,32] within a vehicle. Topology generation and optimization seeks to adjust the configuration or architecture of the system components [33,34]. These methods can be combined to simultaneously optimize both the sizing and topology of the plant to maximize performance. The aircraft design literature also explores sizing and topology optimization, often with an emphasis on structural design [35–37].

*Advancements in control co-design.* The optimization of plant and controller features, also referred to as control co-design (CCD), seeks to maximize closed-loop plant performance. Sequential optimization is used in practice often, where the plant design is first optimized, and then the controller design is then optimized for that plant [38]. This can be extended to an alternating or iterative procedure [39]. However, unless the plant design and controller design problems are separable, the optimal plant and controller variable values must be identified using simultaneous or nested optimization strategies [40]. CCD methods have been applied to automotive systems, with a focus on HEV powertrains with electro-mechanical dynamics. Studies include optimization of component sizes alongside the supervisory controller parameters [41,42], and optimization of the gear ratios and topology with a nested control problem [43].

The maturity of the electrified mobility literature provides studies on plant, controller, and CCD optimization strategies. However, there are three notable gaps in this literature. First, while there is a shift towards inclusion of thermal dynamics in the model, most studies emphasize electro-mechanical dynamics of these systems. Limiting the modeling of thermal dynamics in these systems does not address the integration issue of electrical components whose performance is highly dependent on temperature. Second, and as a result of the first gap, the application of CCD optimization frameworks has not been explored in detail for electrified mobility systems with thermal, electrical, and mechanical dynamics. Third, CCD frameworks center on nested control problems [43], and occasionally tune controller parameters as well [44]. However, studies of CCD with MPC have yet to be explored in detail for electrified mobility, being limited to chemical plant applications [45–47]. To apply CCD frameworks to electrified systems, the controller strategies embedded within CCD frameworks should be inclusive to the MPC strategies used within these systems.

This article addresses these gaps through three contributions to the literature:
1. A thermo-electro-mechanical model of a power-split HEV powertrain is developed. This model captures the dynamics of the electrical and mechanical powertrain components, as well as the thermal management subsystem. The graph-based structure of the model lends itself to be easily implemented in design optimization frameworks, as well as development of MPC algorithms.
2. Simultaneous plant and controller optimization is applied to systems with thermal, electrical and mechanical dynamic elements, using the HEV model as the candidate system. The simultaneous optimization of the plant and controller design variables is compared against plant feature optimization and sequential optimization.
3. MPC is embedded into the plant and controller optimization framework, reflecting the advanced control strategies appropriate for systems with multi-domain dynamics. This allows for the optimization of MPC parameters such as the time step, preview, control input limits, and state tracking gains. This article will focus on centralized MPC, leaving extensions to hierarchical MPC for future studies.

To develop these contributions, Section II reviews the graph-based modeling techniques used to create the HEV model. It also reviews a recently developed design optimization framework for connecting graph-based models to plant and controller design optimization. Section III presents the formulation of the MPC problem, and Section IV provides an overview of the thermo-electro-mechanical HEV model. In Section V, the design optimization problem is formulated and the model under MPC control is validated for a base design. Section VI presents the optimization results from (i) only optimization of the HEV plant design, (ii) sequential optimization of the plant and controller, and (iii) simultaneous optimization of the plant and controller. The article ends with a summary of the contributions and insights from the optimization studies.

**II. GRAPH-BASED MODELING**

This section reviews the techniques utilized to model multi-domain dynamic systems and perform optimization. Component and system dynamics are modeled using a conservation-based, graphical modeling technique. Optimization is performed using a four step design optimization framework based on modifying the elements in the graph-based models using design variables.

**A. Graph-Based Modeling**

The components and systems in this article are modeled using a modular graph-based modeling approach [48–50]. Figure 1 presents the elements of a graph-based model. A vertex represents an energy storage element, and is described by a storage capacitance $C_i$ and a state $x_i$. An external vertex describes a state $x_k^s$ with dynamics external to the model. An edge represents a power flow $y_j$, and is described by the nonlinear function $f_j$:

$$y_j = f_j\left(x_j^{head}, x_j^{tail}, u_j\right) \quad (1)$$

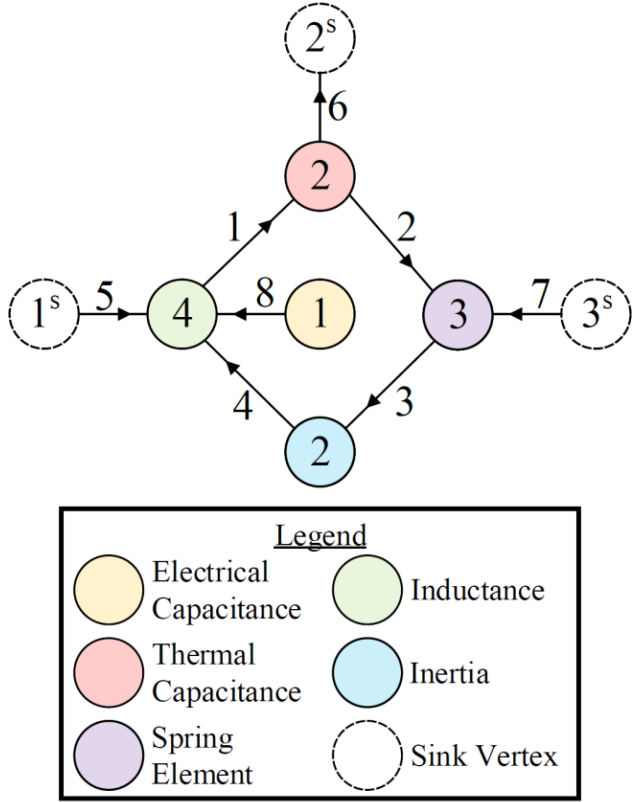

**Figure 1: Graph-based model with the primary elements.**

**Table 1. Vertex types.**

| Vertex Type | Storage Capacitance $C_i$ |
|---|---|
| 1 | $c_i$ |
| 2 | $c_i x_i$ |
| 3 | $c_i(x_i)$ |

**Table 2. Edge types.**

| Edge Type | Power Flow $y_j$ |
|---|---|
| 1 | $\alpha_j x_j^{tail}$ |
| 2 | $\alpha_j x_j^{head}$ |
| 3 | $\alpha_j x_j^{tail} x_j^{head}$ |
| 4 | $\alpha_j \left(x_j^{tail}\right)^2$ |
| 5 | $\alpha_j \left(x_j^{tail} - x_j^{head}\right)$ |
| 6 | $\alpha_j \left(x_j^{tail}\right)^3$ |
| 7 | $\alpha_j u_j x_j^{tail}$ |
| 8 | $\alpha_j u_j x_j^{tail} x_j^{head}$ |

In Eq. (1), $x_j^{head}$ is the state of the vertex at the head of the edge, $x_j^{tail}$ is the state of the vertex at the tail of the edge, and $u_j$ is an input. Note that all edges are not required to have an input. A graph with $i = 1, \ldots, N_v$ vertices, $j = 1, \ldots, N_e$ edges, and $k = 1, \ldots, N_t$ sink vertices is described by:

$$C\dot{x} = -\bar{M}y \quad (2)$$

In Eq. (2), $C \in \mathbb{R}^{N_v \times N_v}$ is the diagonal matrix of storage capacitances, $x \in \mathbb{R}^{N_v}$ is the vector of states, $y \in \mathbb{R}^{N_e}$ is the vector of power flows, and $\bar{M} \in \mathbb{R}^{N_v \times N_e}$ maps the power flows to the state vector. An entry of $\bar{M}$ is $+1$ if a vertex is the tail of an edge, $-1$ if a vertex is a head of an edge, and 0 otherwise. Note that the models presented in this article do not feature source edge elements in the graph.

The models presented in this article contain three types of vertices, described in Table 1. Type 1 vertices often describe thermal capacitances. Type 2 vertices typically describe inductances, electrical capacitances, inertias, and mechanical spring elements. Type 3 vertices are reserved for storage capacitances relating to nonlinear behaviors, such as those found in batteries. Similarly, Table 2 presents the eight categories of edges used in the models of this work. Note that any of these edges can have a nonlinear expression for the parameter $\alpha_j$.

### B. Multi-Domain Design Optimization Framework

Plant and controller optimization is performed using a design optimization framework for multi-domain conservation-based systems [51]. Starting with a graph-based model described by Eq. (2), optimization is performed using four steps:

1. The graph-based model is augmented with design matrices to describe sizing and topology changes, yielding Eq. (3). Sizing of elements is controlled by continuous plant design variables $\boldsymbol{\theta}_c \in \mathbb{R}^{N_c}$, and topology changes are controlled by discrete plant design variables $\mathbf{z} \in \mathbb{R}^{N_z}$. The design matrix $\boldsymbol{\Psi}_c \in \mathbb{R}^{N_v \times N_v}$ ($\boldsymbol{\Phi}_c \in \mathbb{R}^{N_v \times N_v}$) is a diagonal matrix that maps the continuous (discrete) plant design variables to modifications of the storage capacitances. Similarly, the matrix $\boldsymbol{\Psi} \in \mathbb{R}^{N_e \times N_e}$ ($\boldsymbol{\Phi} \in \mathbb{R}^{N_e \times N_e}$) is a diagonal matrix that maps the continuous (discrete) variables to modifications of the power flows.

$$\boldsymbol{\Psi}_c C\dot{x} = -\bar{M}\boldsymbol{\Psi}\boldsymbol{\Phi}y \quad (3)$$

2. The objective function is defined for optimization. Eq. (4) defines the aggregate objective function $J_{tot}$ for this work, with 4 individual objectives:

$$J_{tot} = w_{st}J_{st}^m + w_{sc}J_{sc}^m + w_{en}J_{en}^m + w_{size}J_{size}^m \quad (4)$$

Eq. (5) defines the state tracking error objective, $J_{st}$, with weight $w_{st}$. The objective is evaluated at discrete times $j = 1, \ldots, N_{final}$ with timestep $\Delta t_e$ and final time $t_{final} = N_{final}\Delta t_e$. The vector $r \in \mathbb{R}^{N_v}$ contains the reference states, and the matrix $\mathbf{\Lambda}_x \in \mathbb{R}^{N_v \times N_v}$ internally weighs the terms, with $\|x\|_{\Lambda_x}^2 = x^T \Lambda_x x$.

$$J_{st} = 10^{-7} \sum_{j=1}^{N_{final}} \|x_j - r_j\|_{\Lambda_x}^2 \quad (5)$$

Eq. (6) defines the state constraint violations objective, $J_{sc}$, with weight $w_{sc}$, evaluated at the same times as $J_{st}$. The slack variables $s \in \mathbb{R}^{N_v}$ are equal to the maximum of $[|x_i - x_{\min,i}|, |x_i - x_{\max,i}|]$ for state $x_i$, with $x_{\min,i}$ and $x_{\max,i}$ as the minimum and maximum allowable state values, respectively. The matrix $\mathbf{\Lambda}_s \in \mathbb{R}^{N_v \times N_v}$ internally weighs the terms.

$$J_{sc} = 10^{-10} \sum_{j=1}^{N_{final}} \|s_j\|_{\Lambda_s}^2 \quad (6)$$

Eq. (7) defines the energy utilization objective, $J_{en}$, with weight $w_{en}$. This function relates to the total energy passing through the edges over the test period, with the integral evaluated using the trapezoidal rule with the same time interval as $J_{st}$. The matrix $\mathbf{\Lambda}_p \in \mathbb{R}^{N_e \times N_e}$ internally weighs the terms, with $\Lambda_{p,j,j}$ as the internal weight for edge $j$.

$$J_{en} = 10^{-7} \sum_{j=1}^{N_e} \left( \int_0^{t_{final}} \Lambda_{p,j,j} y_j(t) \, dt \right) \quad (7)$$

These first three objectives are calculated by evaluating the dynamics of the system under open-loop or closed-loop control. Eq. (8) describes the final objective $J_{size}$, with weight $w_{size}$, to minimize the size of the continuous plant design variables, with internal weights $w_c$.

$$J_{size} = 10^{-1} w_c \theta_c \quad (8)$$

Note that each of these objectives are internally normalized by a term (e.g., $10^{-7}$, $10^{-10}$, $10^{-1}$). These four objectives are chosen to directly capture important performance, efficiency, and sustainability metrics for automotive applications. For identification of optimal designs for nonconvex problems, the integer $m = 4$ is used for compromise programming when optimizing.

3. The constraints for the optimization problem are defined. The plant design variables $\theta_c$ and $z$ are bounded between minimum values $\underline{\theta_c}$ and $\underline{z}$, and maximum values $\overline{\theta_c}$ and $\overline{z}$. In addition, to optimize controller parameters, the controller design variables $\phi \in \mathbb{R}^{N_\phi}$ are defined [52]. These have minimum and maximum bounds $\underline{\phi}$ and $\overline{\phi}$, respectively. Eq. (9) defines the constraints.

$$\begin{aligned} \underline{\theta_c} &\leq \theta_c \leq \overline{\theta_c} \\ \underline{z} &\leq z \leq \overline{z} \\ \underline{\phi} &\leq \phi \leq \overline{\phi} \end{aligned} \quad (9)$$

4. The optimization problem is formulated using the previous three steps and solved. The shooting method is used by simulating the dynamics of Eq. (3) and evaluating the objective function of Eq. (4). Design variables are updated to ensure the constraints of Eq. (9) are met.

### III. MPC FORMULATION

This section reviews the formulation of the centralized MPC problem [27,53]. This formulation is used to control the HEV powertrain energy dynamics within the design optimization process. The prediction model for MPC is determined by linearizing the augmented graph-based model of Eq. (3) about the most recent operating states and inputs of the system at a given time index $p$. This creates a set of differential algebraic equations (DAEs), which are discretized temporarily using the forward Euler method and timestep $\Delta t$. Eq. (10) presents the linearized, discretized model equations, with $\mathbf{C}_p \in \mathbb{R}^{N_v \times N_v}$ as a diagonal matrix akin to the capacitance matrix, $\mathbf{A}_p \in \mathbb{R}^{N_v \times N_v}$ similar to the state matrix, and $\mathbf{B}_p \in \mathbb{R}^{N_v \times N_u}$ similar to the input matrix. The column vector $u \in \mathbb{R}^{N_u}$ contains $N_u$ inputs. The matrix $\mathbf{V}_p \in \mathbb{R}^{N_v \times N_t}$ describes how external states impact the dynamics, $x^s \in \mathbb{R}^{N_t}$ are the external states, and $\mathbf{W}_p \in \mathbb{R}^{N_v}$ describes constant terms created from linearization about a non-equilibrium point.

$$\mathbf{C}_p x_{k+1} = (\mathbf{C}_p + \Delta t \mathbf{A}_p) x_k + \Delta t \mathbf{B}_p u_k + \Delta t \mathbf{V}_p x_k^s + \Delta t \mathbf{W}_p \quad (10)$$

To stabilize the discrete model for a given timestep, model reduction is performed by setting select entries of $\mathbf{C}$ to 0.

Eq. (11) defines the optimization problem solved for each time the controller is called, determining inputs $[u_{k+1|k}, \ldots, u_{k+N_p|k}]$. The first constraint of Eq. (11) defines the state dynamics. The second through fourth constraints describe the conditions for the slack variables, and the fifth constraint bounds the inputs between $u_{\min}$ and $u_{\max}$. The sixth constraint is to keep inputs within a range of the previous input $u_{k|k} = u_{k+1|k-1}$, using variable $0 \leq \varepsilon \leq 1$. The seventh constraint fixes certain inputs to be equal to state values, using matrices $Z_1 \in \mathbb{R}^{N_u \times N_u}$ and $Z_2 \in \mathbb{R}^{N_u \times N_v}$. This is used to relate pump and fan speeds to mass flow rates. The eight constraint provides the final value for the algebraic states $x^a$ at $k + N_p + 1$, which cannot be calculated without knowledge of inputs at this time index. The final constraint sets the initial conditions for the dynamic states $x^d$, provided from the plant dynamics when the controller is called.

$$J_k^* = \min_{\overline{u}_k} J_k$$

$$s.t.\ \forall j \in [1, N_p] \quad C_k x_{k+j+1|k} = (C_k + \Delta t A_k) x_{k+j|k}$$
$$+ \Delta t B_k u_{k+j|k} + \Delta t V_k x_{k+j|k}^s$$
$$+ \Delta t W_k$$
$$x_{\min} - s_{k+j|k} \leq x_{k+j+1|k}$$
$$x_{k+j+1|k} \leq x_{\max} + s_{k+j|k} \quad (11)$$
$$s_{k+j|k} \geq 0$$
$$u_{\min} \leq u_{k+j|k} \leq u_{\max}$$
$$(1-\varepsilon)u_{k|k} \leq u_{k+j|k} \leq (1+\varepsilon)u_{k|k}$$
$$Z_1 u_{k+j|k} = Z_2 x_{k+j|k}$$
$$x_{k+N_p+1|k}^a = x_{k+N_p|k}^a$$
$$x_{k+1|k}^d = x_k^d$$

Eq. (12) defines the objective function $J_k$ at time index $k$, which is an aggregate of two terms. The first term represents state tracking error with weighting matrix $\Lambda_{x,c}$, while the second represents state constraints with weighting matrix $\Lambda_{s,c}$. These functions are similar to Eq. (5) and (6), but are evaluated only over the horizon of $N_p$ steps. The predicted state vector for calling the controller at time index $k$ is $x_{k+j|k}$, for $j = 1, \ldots, N_p$.

$$J_k = \sum_{j=1}^{N_p+1} \|x_{k+j|k} - r_{k+j|k}\|_{\Lambda_{x,c}}^2 + \sum_{j=1}^{N_p} \|s_{k+j|k}\|_{\Lambda_{s,c}}^2 \quad (12)$$

## IV. HEV MODEL

This section presents a thermo-electro-mechanical model of an HEV powertrain with cooling. To retain clarity for this article, the full description of the model is provided in the supplementary document for this article. The HEV model is built by combining graph-based models of individual components. The components are separated into four categories. The first category contains components that operate primarily with electrical dynamics. This includes batteries, buses, and power electronics such as inverters, rectifiers, and DC-DC converters. The second category contains components that operate primarily with mechanical dynamics, including the motors and generators, pumps and fans, planetary gears and transmissions, and engines. The third category contains components with thermal dynamics, such as vapor compression systems (VCS) and cooling loops. The fourth category contains connector graphs that enable the combination of separate component graphs, such as gearboxes, radiators, and virtual inductors to maintain proper graph-based modeling format [50]. Multiple versions of component graphs are used within the model (e.g., there are multiple DC/DC converters). The numbered component graphs are as follows:

1. Battery
2. High voltage bus
3. Inverter
4. Motor
5. Transmission
6. Wheels and chassis
7. Converter #1
8. Low voltage bus
9. Air cooling path
10. Power electronics liquid cooling loop
11. Radiator #1
12. Thermal combiner
13. Virtual inductor #1
14. Virtual inductor #2
15. Converter #2
16. Converter #3
17. Pump #1
18. Fan #1
19. VCS
20. Converter #4
21. Virtual inductor #3
22. VCS liquid cooling loop
23. Battery liquid cooling loop
24. Pump #2
25. Converter #5
26. Virtual inductor #4
27. Cabin air path
28. Cabin
29. Fan #2
30. Converter #6
31. Virtual inductor #5
32. Transaxle liquid cooling loop
33. Radiator #2
34. Pump #3
35. Converter #7
36. Virtual inductor #6
37. Rectifier
38. Generator
39. Gearbox #1
40. Gearbox #2
41. Engine
42. Planetary gear

Figure 2 and Figure 3 present the configuration of the full thermo-electro-mechanical model of the power-split HEV powertrain with cooling. Component models are interconnected to produce the full model, using a graphical composition method [54]. Graph interconnections are defined primarily using edge equivalencies: two edges, each from different graphs, are defined to be combined. The head and tail vertices of the edge from the first graph are matched to the tail and head vertices, respectively, of the second graph, defining vertex equivalencies. As the two combined edges can have different parameters or be of different types, one graph's edge is defined to be the dominant one used in the combined model. The supplementary material lists all component graphs used and defines the edge equivalencies and dominant edges to create the full HEV model.

## V. PROBLEM SETUP

This section defines the design optimization problem for the HEV powertrain with cooling. Six plant design variables and three controller design variables are selected. Using baseline design variable values, the plant dynamics operating under the MPC algorithm are validated.

### A. Problem Formulation

Six plant design variables are defined to optimize for the HEV system:

1. $\theta_{c,1}$ is size of the heat exchanger near the battery.
2. $\theta_{c,2}$ is size of the heat exchanger near the power electronics box.
3. $\theta_{c,3}$ is size of the heat exchanger near the motor.
4. $\theta_{c,4}$ is size of the heat exchanger near the generator.
5. $\theta_{c,5}$ is size of the heat exchanger near the planetary gear box.
6. $\theta_{c,6}$ is the number of battery cells in parallel within the battery pack. This also scales the mass of the vehicle to account for larger battery pack sizes.

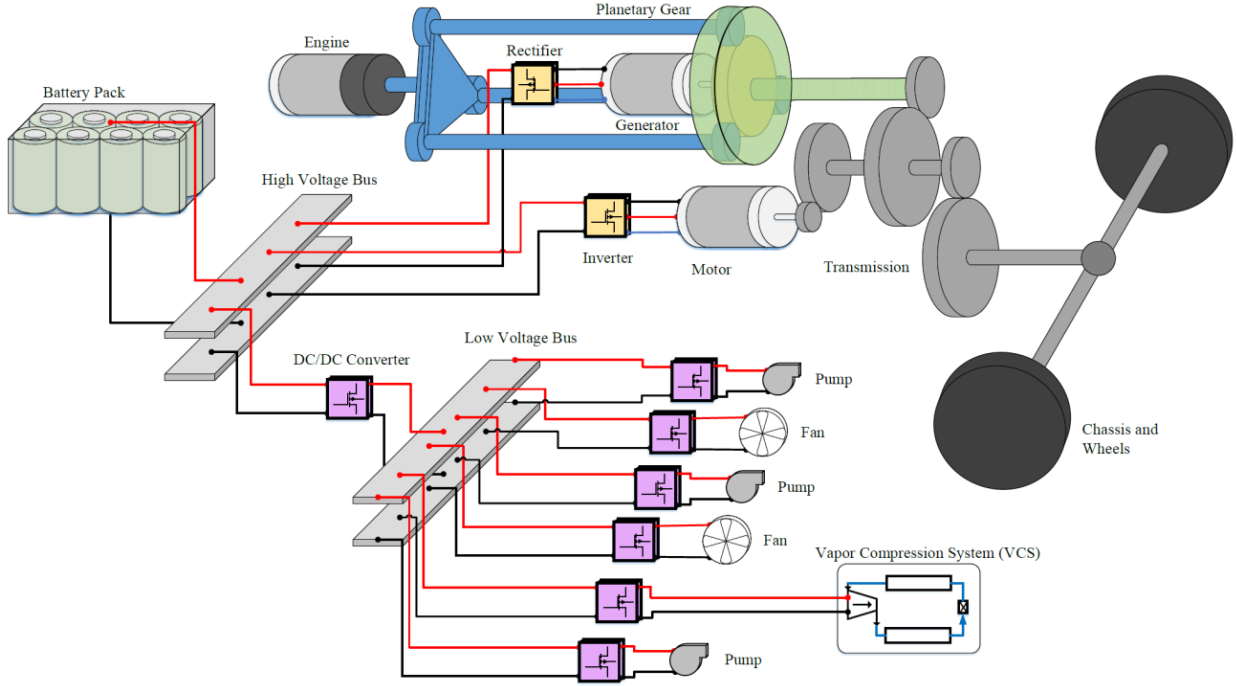

**Figure 2: Electro-mechanical configuration of the HEV powertrain.**

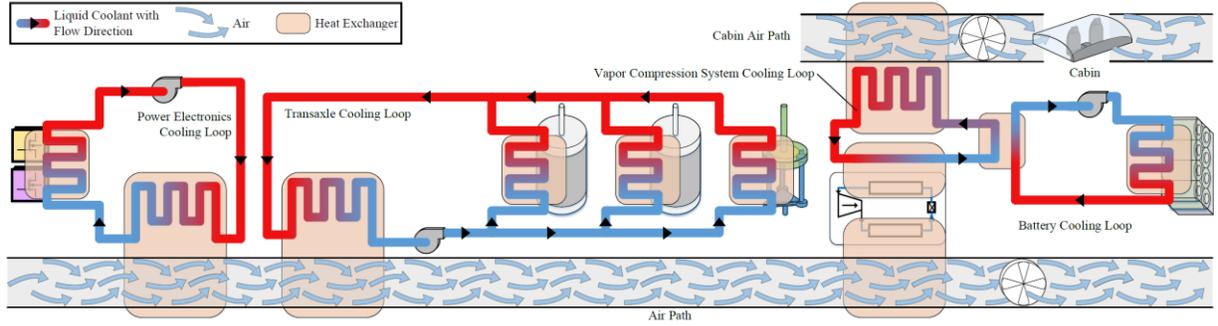

**Figure 3: Thermal cooling system for the HEV.**

A detailed description of how these relate to $\boldsymbol{\Psi_c}$ and $\boldsymbol{\Psi}$ is provided in the supplementary material. While the design framework is inclusive to topology optimization, such as selection of air vs. liquid cooling [51], it is not performed in this work. As such, there are no discrete plant design variables, and $\boldsymbol{\Phi_c}$ and $\boldsymbol{\Phi}$ are both equal to the identity matrix. Eq. (13)-(14) present the constraints for the plant design variables.

$$\underline{\boldsymbol{\theta_c}} = [0.1, 0.1, 0.1, 0.1, 0.1, 1]^T \quad (13)$$

$$\overline{\boldsymbol{\theta_c}} = [100, 100, 100, 100, 100, 10]^T \quad (14)$$

Three controller design variables are defined, with discrete options for each:

1. $\phi_1 = \Delta t$ is the timestep for the controller. It is constrained to values of 0.5, 1, 2, and 3.
2. $\phi_2$ is the allowable perturbation of inputs from the last input, e.g. $\phi_2 = \varepsilon$. This controller design variable is constrained to 0.2, 0.25, 0.3, 0.35, and 0.4.
3. $\phi_3$ is the internal weight within $\boldsymbol{\Lambda_{x,c}}$ relating to velocity tracking. It is constrained to the following values: 10, 25, 50, 100 to 1000 in increments of 100, 1500, 2000, 3000, 5000, 10000, 50000, and 100000.

The design objective functions are the same as those presented in Eq. (4)-(8), with 0.5 used for $w_{st}$, $w_{sc}$, $w_{en}$, and $w_{size}$. State tracking error focuses on tracking vehicle velocity with a weighting of 100. The battery surface temperature is constrained between 20°C and 40°C. The motor and generator temperatures are constrained between 0°C and 80°C, and the power electronics temperatures are constrained between 10°C and 110°C. All temperature constraints have an internal weight of $1.4 \times 10^3$. The battery state of charge (SOC) is constrained between 0.3 and 0.7, with an internal weight of $1.4 \times 10^5$. Total energy utilization is determined from battery edges 4-6 and 8, and engine edge 1, yielding the total energy used while driving. For the sizing objective, $w_c$ factors the heat exchanger sizes equally with internal values of 1, and an internal value of 0 for the battery pack size.

The controller tracks the desired velocity with internal weight $\phi_3$, and contains the same constraints and internal weights as the design optimization problem. The duty cycles for the power electronics make up a majority of the inputs for the system. The duty cycle for the converter between the high and low voltage buses is limited between 0.19 and 0.2, while the duty cycle input for the converter connected to the VCS is limited between 0.09 and 0.1. The duty cycle for converters powering fans and pumps are limited to 0.04 to 0.05 and 0 to 0.3, respectively. The inverter duty cycle, rectifier duty cycle, engine throttle, and brake command are all limited between 0 and 1. The input to each cooling system and air path is the mass flow rate, which is dependent on the speed of the pump or fan connected to it. The controller operates with a horizon $N_p = 4$ steps. In addition, the first two inputs provided from the controller are used before calling the controller again.

### B. Model and Controller Validation

The model and centralized MPC algorithm are validated using a base set of plant and controller design variables. The UDDS drive cycle is used to define a desired vehicle velocity. Ambient temperature profiles are at a constant 25°C, with all other external state profiles set to zero. In this example, $\theta_{c,1} = \cdots = \theta_{c,5} = 100$, $\theta_{c,6} = 3$, $\phi_1 = 1$, $\phi_2 = 0.25$, and $\phi_3 = 100$. The MATLAB `ode23tb` solver is used to solve the system, with analytical solutions provided for the algebraic state equations.

Figure 4 presents the state trajectories for the battery SOC, vehicle velocity, battery temperature, and planetary gear temperature. The plant configuration and controller are able to track the desired vehicle velocity, with an average error of 1.20 mph, with error defined as the absolute value of the difference of the reference and actual. State constraint violations are minimal, with maximum violations of 0.0084 for battery SOC, 0°C for battery temperature, and 0.14°C for the planetary gear. The motor, generator, and power electronics all remain within their temperature constraints.

### VI. OPTIMIZATION RESULTS

This section presents optimization results for the thermo-electro-mechanical HEV model with MPC. The plant and controller design variables are optimized for the UDDS drive cycle. Three studies are performed using a genetic algorithm (GA) to optimize, with a population size of 10X the number of design variables being adjusted. The first study identifies the six plant design variable values for fixed controller design variable values. The second two studies perform two versions of CCD, including sequential optimization and simultaneous optimization of the plant and controller design variables. Results are presented to show the improvements in performance of the plant with controller.

### A. Plant Optimization with the UDDS Drive Cycle

Optimization using the UDDS drive cycle is first performed by optimizing the plant design variables with fixed controller design variables, using the values of $\phi_1 = 1$, $\phi_2 = 0.25$, and $\phi_3 = 100$. Optimization is performed using the MATLAB `ga` function with a population size of 60 and default settings. A computing cluster node with 192 GB of memory and 36 cores is used to run the optimization process in parallel. A total of 30 populations were generated, with 1800 function evaluations total. The set of design variables with the lowest total objective function value was generated in the 11[th] population.

Table 3 presents the design variables and objective function values for the designs with the lowest objective function values from the first population (initial design) and the final population (optimized plant design). Figure 5 presents the SOC, velocity, battery temperature, and planetary gear temperature state trajectories for both of these system designs. The average velocity tracking error is 1.41 mph for the initial design, and 1.27 mph for the optimized plant design. Neither design violates state constraints. There is a 58.2% reduction in the size of the heat exchangers after optimizing for the plant design variables. The total objective function is reduced by 38.6%.

### B. Plant and Controller Optimization

To quantify the advantage conferred by control co-design, we compare our results of plant-only optimization against sequential and simultaneous optimization of the plant and controller design variables values. Sequential optimization uses a GA to identify the controller design variables values. The plant design variables fixed, using the values identified from the optimized plant of Section VI.A. Simultaneous optimization describes the study which identifies the values for plant and controller design variables simultaneously using a GA.

For the sequential optimization study, an additional 10 populations of size 30 were evaluated after the optimization of the plant design variables, adding 300 to the earlier 1800 function evaluations. The total number of function evaluations was 2100, and the best controller design variables were generated in the 3[rd] population. For the simultaneous optimization study, the GA was run for 20 generated populations of size 90, for a total of 1800 function evaluations.

Table 4 presents the design variables and objective function values using sequential and simultaneous optimization of the plant and controller design variables (sequential and simultaneous CCD-MPC, respectively). Both reduce the total objective function as compared to the system determined from solely optimizing the plant design variables. However, the total objective function from the simultaneous method is 20% lower as compared to the sequential method. This is primarily due to the reduction the simultaneously identified system provides in state tracking error and heat exchanger sizes. Unlike sequential design, simultaneous CCD is able to optimally exploit the inherent design coupling between the plant and the controller. Figure 6 presents the state trajectories for the system identified using the simultaneous method as compared to the system identified from optimizing the plant only. As compared to the optimized plant design, the simultaneously identified design reduces state tracking error as Figure 6(b) shows, with an average error of 0.99 mph. However, Figure 6(d) indicates that the number of state constraint violations is increased in the planetary gear, reaching a maximum of 7.52°C over the constraint for a brief period of time. This is a trade-off that is determined by the relative weighting of the individual objective functions. The increase in state constraint violations is a relatively small increase in the total objective function, being 0.0376 after factoring in the weight $w_{sc}$. However, by doing so, the total size of the heat exchangers is reduced significantly, dropping the contribution of this to the total objective function by 0.31, an order of magnitude larger impact than the constraint violation.

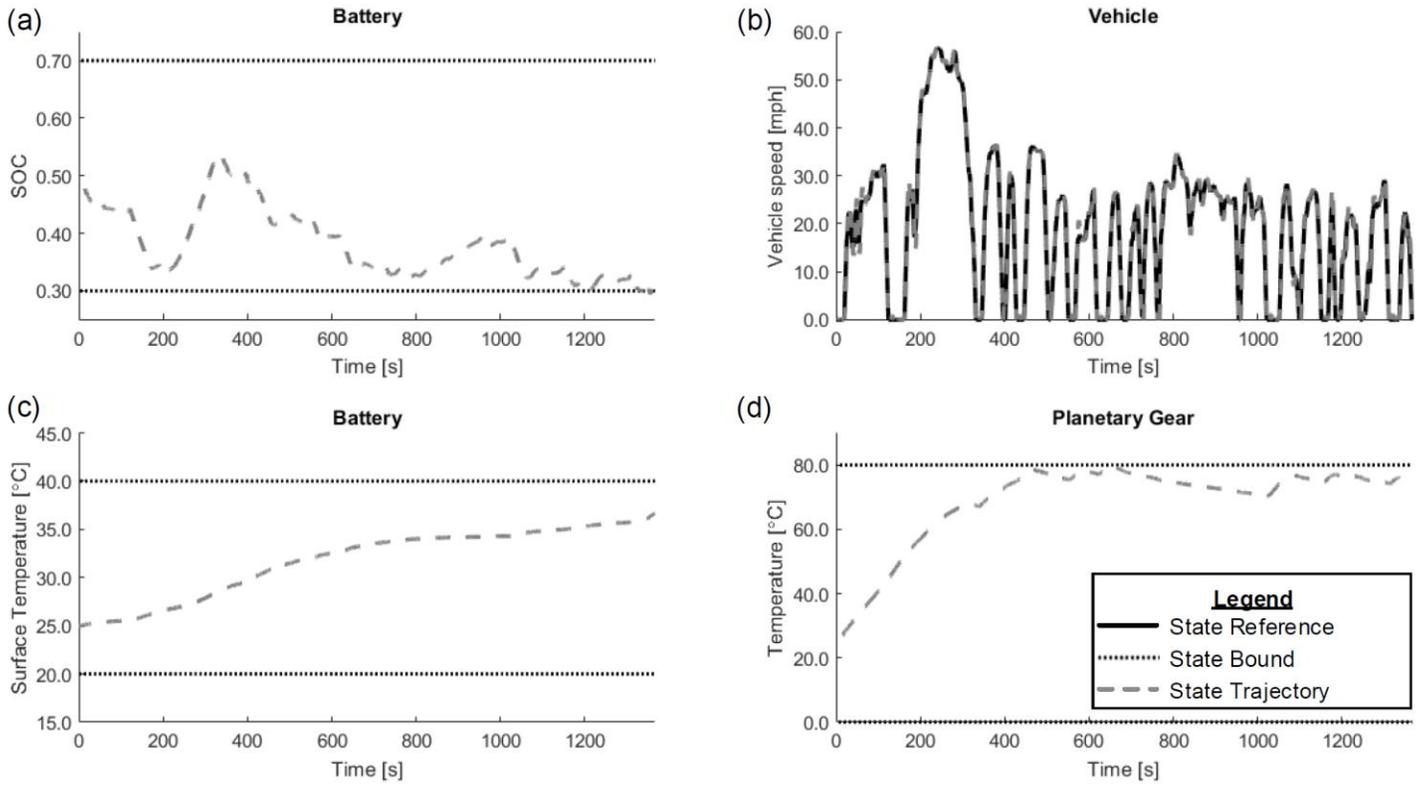

**Figure 4: (a) Battery SOC, (b) velocity, (c) battery temperature, and (d) planetary gear temperature for model and controller validation.**

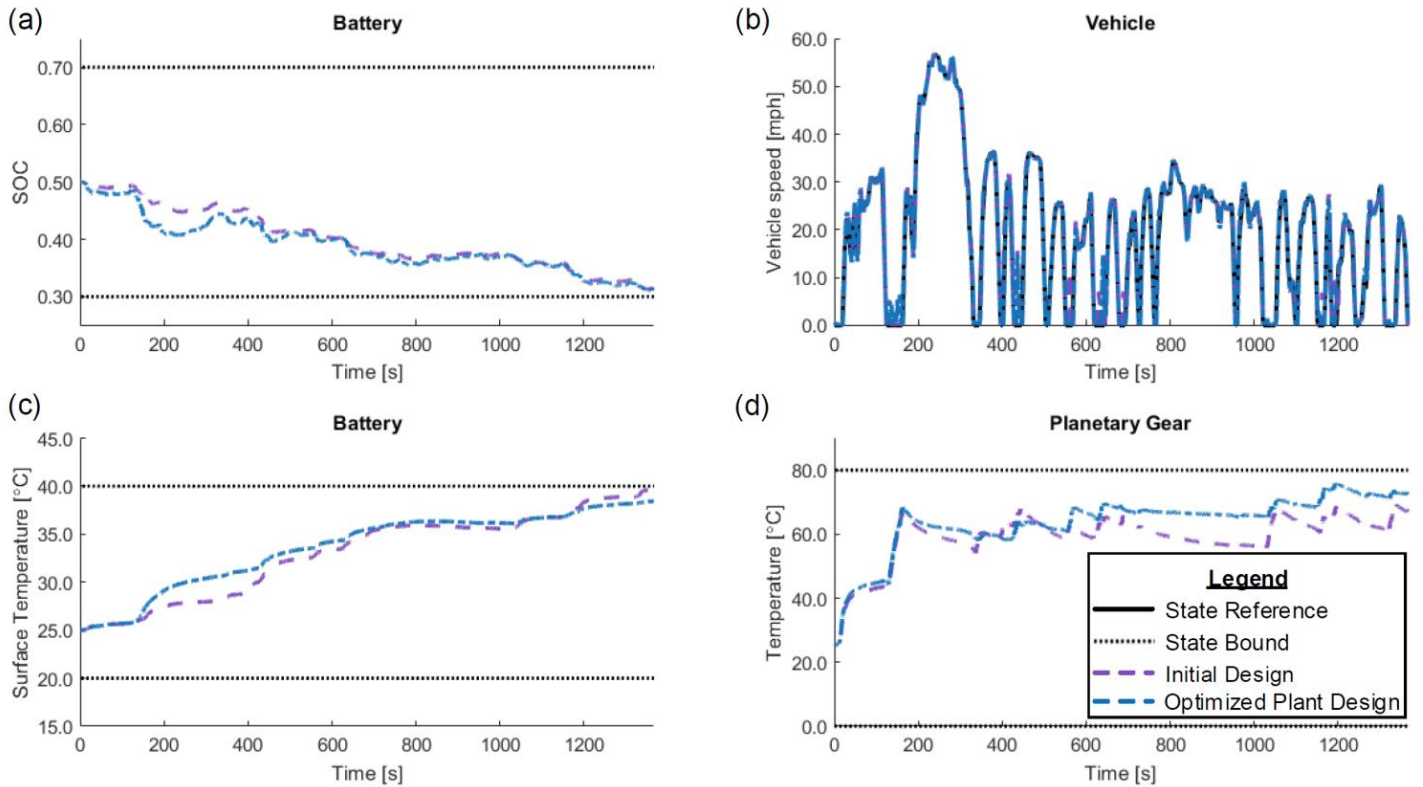

**Figure 5: (a) Battery SOC, (b) velocity, (c) battery temperature, and (d) planetary gear temperature for the best initial design and the optimized plant design.**

Figure 7(a) presents a spider plot of the weighted individual objective functions (e.g., $w_{st}J_{st}$, $w_{sc}J_{sc}$) for each of the system designs. As shown, simultaneous optimization of the plant and controller design variables yields the largest reduction in the objective function, approaching closest to the unreachable utopia point. Sequential optimization of the plant and controller design variables does create improvements as compared to only optimization of the plant design, but not by as much as using simultaneous methods. The trend is reflected in Figure 7(b), which plots the total objective function values ($m = 1$).

Table 5 quantifies how plant optimization, sequential plant and controller optimization, and simultaneous plant and controller optimization reduce the objective functions as compared to the best design from the initial population. Simultaneous optimization reduces state tracking error by 57.4%, while sequential optimization and plant optimization reduce it by 41.3% and 18.7%, respectively. Simultaneous optimization reduces the total heat exchanger size by 62.3%, slightly more than the 58.2% from plant optimization. The trade-offs made are that simultaneous optimization has a small, nonzero increase in state constraint violations (an infinite increase from 0), and a smaller reduction in energy used. However, these trade-offs allow for a 57.3% reduction in the total objective function, as compared to a 38.6% reduction found by only performing plant optimization. This matches expectations: simultaneous optimization of plant and controller design variables yield superior designs as compared to only optimizing the plant, or optimizing the plant and the controller in a sequential fashion. These results indicate that this insight also applies to plants controlled by MPC algorithms.

## VII. CONCLUSIONS

This article explores control co-design (CCD) for power and energy systems with the integration of model predictive control (MPC). While the electrified mobility literature contains studies on plant and controller optimization for electro-mechanical powertrain dynamics, thermal dynamics are generally excluded. As electrical component performance is strongly coupled to temperature, this article expands on previous works by developing a novel thermo-electro-mechanical hybrid electric vehicle (HEV) model suitable for CCD. To utilize a controller recognized for handling thermal and electrical dynamics within these systems, a MPC algorithm is used to control the system dynamics.

**Table 3. Design variables and objective function values for the initial and plant-optimized systems.**

|  | Initial Design | Optimized Plant Design |
|---|---|---|
| $\theta_{c,1}$ | 22.5 | 38.2 |
| $\theta_{c,2}$ | 7.91 | 1.56 |
| $\theta_{c,3}$ | 73.2 | 0.639 |
| $\theta_{c,4}$ | 29.7 | 8.64 |
| $\theta_{c,5}$ | 16.9 | 13.7 |
| $\theta_{c,6}$ | 7.28 | 4.35 |
| $\phi_1$ | 1 | 1 |
| $\phi_2$ | 0.25 | 0.25 |
| $\phi_3$ | 100 | 100 |
| $J_{st}$ | 13.1 | 10.7 |
| $J_{sc}$ | 0 | 0 |
| $J_{en}$ | 1.48 | 1.25 |
| $J_{size}$ | 15.0 | 6.28 |
| $J_{tot}$ | 14.8 | 9.10 |

**Table 4. Design variables and objective function values for the sequential and simultaneous control co-design systems.**

|  | Sequential CCD-MPC Design | Simultaneous CCD-MPC Design |
|---|---|---|
| $\theta_{c,1}$ | 38.2 | 20.2 |
| $\theta_{c,2}$ | 1.56 | 4.41 |
| $\theta_{c,3}$ | 0.639 | 11.6 |
| $\theta_{c,4}$ | 8.64 | 17.5 |
| $\theta_{c,5}$ | 13.7 | 2.89 |
| $\theta_{c,6}$ | 4.35 | 7.76 |
| $\phi_1$ | 1 | 0.5 |
| $\phi_2$ | 0.25 | 0.2 |
| $\phi_3$ | 1000 | 1000 |
| $J_{st}$ | 7.69 | 5.58 |
| $J_{sc}$ | 0 | 0.0752 |
| $J_{en}$ | 1.24 | 1.32 |
| $J_{size}$ | 6.28 | 5.66 |
| $J_{tot}$ | 7.60 | 6.32 |

**Table 5. Reductions in objective function values as compared to the best initial design. Darker shading indicates a relatively greater reduction as compared to the other designs.**

|  | State Tracking Error | State Constraint Violations | Energy Used | Heat Exchanger Size | Total Objective |
|---|---|---|---|---|---|
| Optimized Plant Design | 18.7% reduction | 0.00% reduction | 15.3% reduction | 58.2% reduction | 38.6% reduction |
| Sequential CCD-MPC | 41.3% reduction | 0.00% reduction | 16.5% reduction | 58.2% reduction | 48.6% reduction |
| Simultaneous CCD-MPC | 57.4% reduction | ∞ increase | 11.0% reduction | 62.3% reduction | 57.3% reduction |

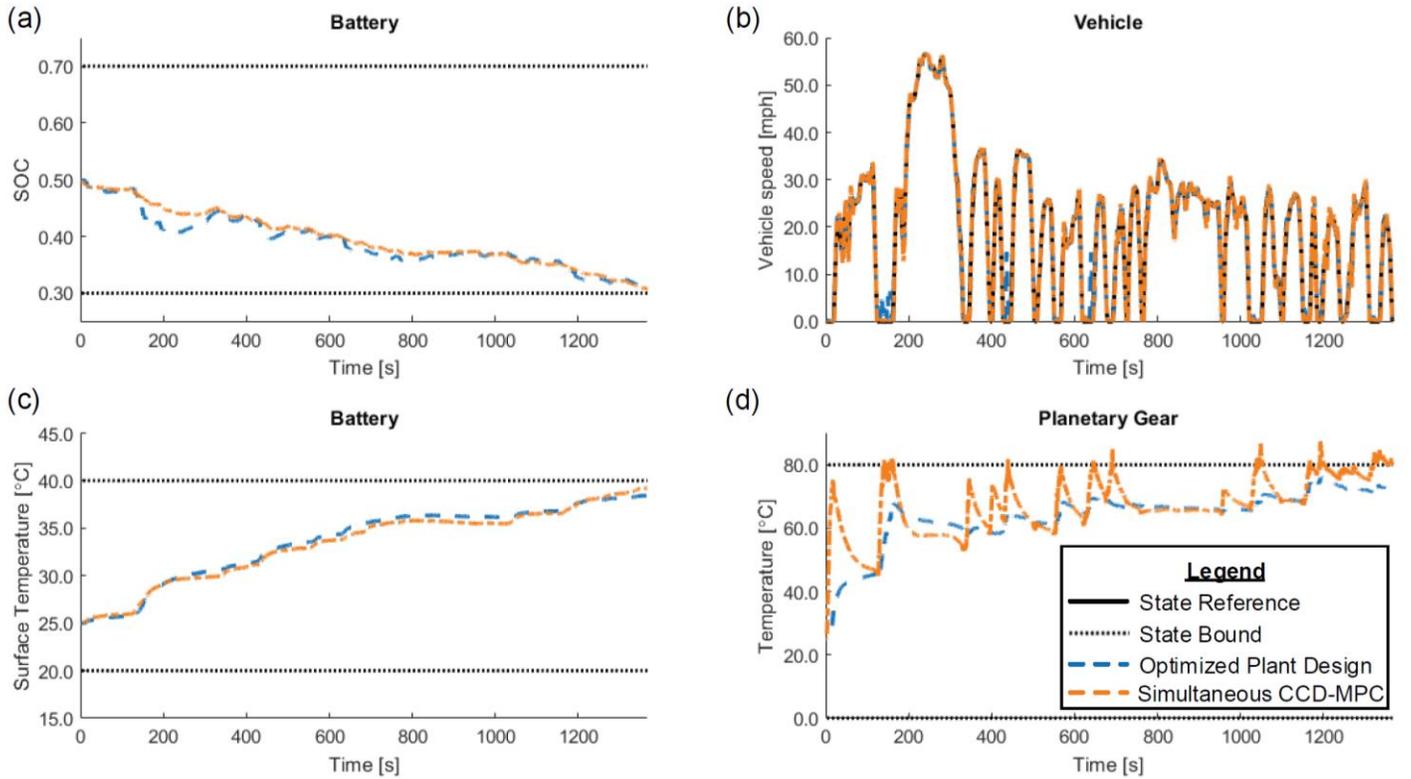

**Figure 6: (a) Battery SOC, (b) velocity, (c) battery temperature, and (d) planetary gear temperature for the optimized plant design and the design identified from simultaneous control co-design.**

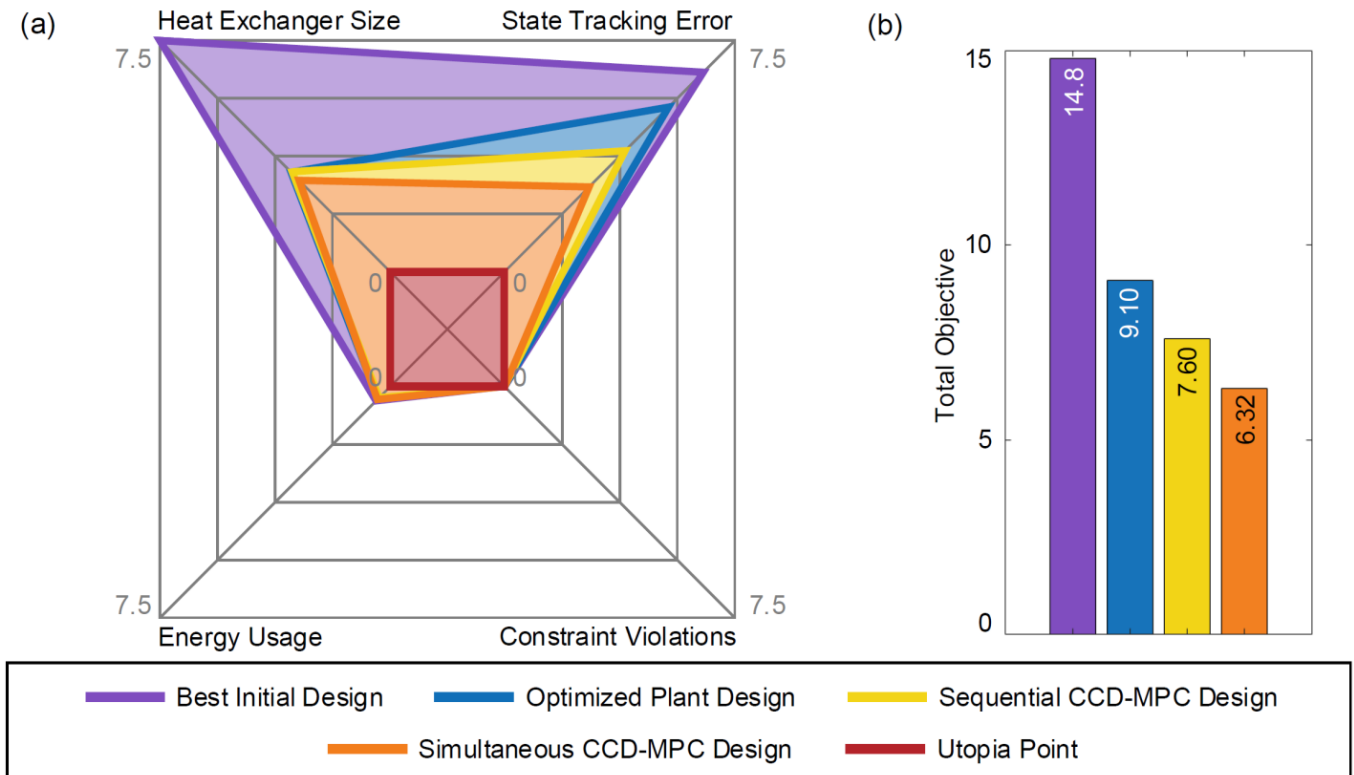

**Figure 7: Comparison of (a) the weighted individual objective function values and (b) the total objective function values for each design identified using different optimization strategies.**

Case studies optimize plant design variables, including heat exchanger and battery sizes, as well as controller design variables such as the MPC timestep, input limits, and MPC gains for tracking. Four individual objective functions, including heat exchanger size, state tracking error, constraint violations, and energy used, make up the total objective. Results show that optimization of the plant with a baseline MPC algorithm reduces heat exchanger size by 58.2% and state tracking error by 18.7% from an initial design. Optimization of both plant and controller design variables, either sequentially or simultaneously, further reduces the total objective. Simultaneous optimization provides the greatest reductions, with a 62.3% reduction in heat exchanger size and a 57.4% reduction in state tracking error, with a trade-off of minimally more state constraint violations. The results indicate that embedding MPC into CCD frameworks can yield significant improvements in electrified vehicle performance. Future work will build on these results and integrate additional controller features and MPC parameters into the design optimization process.

## ACKNOWLEDGMENTS

This material is based upon work supported by the National Science Foundation Engineering Research Center for Power Optimization of Electro Thermal Systems (POETS) with cooperative agreement EEC-1449548. The authors thank Christopher T. Aksland for his valuable feedback during the development of the vehicle model.

# SUPPLEMENTARY MATERIAL: A HYBRID-ELECTRIC VEHICLE MODEL WITH THERMO-ELECTRO-MECHANICAL DYNAMICS

**Donald J. Docimo**
Department of Mechanical Engineering, Texas Tech University
Lubbock, TX, U.S.A.
donald.docimo@ttu.edu

**Ziliang Kang**
Department of Aerospace Engineering, University of Illinois at Urbana-Champaign
Urbana, IL, U.S.A.
kang134@illinois.edu

**Kai A. James**
Department of Aerospace Engineering, University of Illinois at Urbana-Champaign
Urbana, IL, U.S.A.
kaijames@illinois.edu

**Andrew G. Alleyne**
Department of Mechanical Science and Engineering, University of Illinois at Urbana-Champaign
Urbana, IL, U.S.A.
alleyne@illinois.edu

## I. OVERVIEW

This document outlines a thermo-electro-mechanical model for a hybrid electric vehicle (HEV) powertrain with cooling. First, the graph-based component models are detailed and parameterized. Second, a description is provided for how these component models are interconnected to create the full HEV model. Finally, plant design variables of interest are described.

## II. GRAPH-BASED COMPONENT MODELS

This section presents the graph-based models for the components used to build a full system. The components are separated into four categories. The first category contains components that operate primarily with electrical dynamics. The second category contains components that operate primarily with mechanical dynamics. The third category contains components with thermal dynamics. The fourth category contains connector graphs that enable the combination of separate component graphs.

### A. Electrical Components

Component models of primarily electrical dynamics include batteries, buses, and power electronics such as inverters, rectifiers, and DC-DC converters. Table 1 and Table 2 present descriptions of the vertices and edges, respectively, for each graph presented in Figure 1. Note that rectifiers are identical to inverters and converters, but with component edges 1, 2, 4, and 6 in the reverse direction. The battery parameters are drawn from scaling experimentally determined parameters [1] for 76 cells in series, with the open-circuit voltage of the battery, $V_{oc}$, as a function of SOC as shown in Figure 2.

### B. Mechanical Components

Component models of primarily mechanical dynamics include motors, generators, pumps, fans, engines, planetary gears, transmissions, and chasses with wheels. Table 3 and Table 4 present descriptions of the vertices and edges, respectively, and Figure 3 presents these graphs. Similar to inverters and converters, the motors, fans, and pumps are all identical in structure, but with different parameter values. Generators are also identical to motors in structure, but with component edges 1, 2, and 5 in the reverse direction. Note that the mass of the vehicle is scaled by the values of the plant design variables. The input for the engine is the throttle command, and the input for the chassis and wheels component is the brake force.

Fan and pump power ratings are drawn from [2] to calculate equivalent edge and capacitance parameter values. The torque produced from the engine, $\tau_e$, is a function of the input throttle and the angular velocity of the engine, and is plotted in Figure 4. The engine inertia, used to get the engine's storage capacitance, is provided in [3]. There are five nonlinear parameters that impact the dynamics of the wheels and chassis model. The first, $f_{w,1}$, represents aerodynamic drag acting on the vehicle, as given by Eq. (1). The function $f_{sig}$ is an offset sigmoid function, provided by Eq. (2).

$$f_{w,1} = 0.0105 f_{sig}(x^{tail}) \quad (1)$$

$$f_{sig}(x) = \frac{2}{1 + e^{-x}} - 1 \quad (2)$$

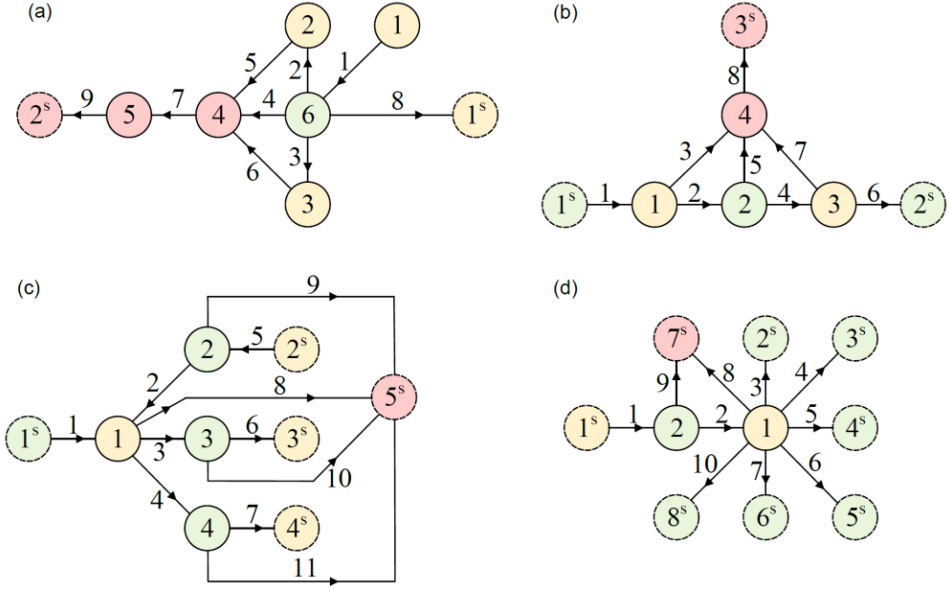

**Figure 1:** Graph-based models for (a) a battery, (b) an inverter, rectifier, or DC-DC converter, (c) a high-voltage bus, and (d) a low-voltage bus.

Table 1. Electrical component vertices.

| | Vertex Number | State | Vertex Type | Parameter Value $c_i$ |
|---|---|---|---|---|
| Battery | 1 | State of Charge (SOC) | 3 | $[1.368*10^6 \text{ C}]*V_{oc}$ |
| | 2 | Voltage 1 | 2 | $3.35*10^5$ F |
| | 3 | Voltage 2 | 2 | $9.21*10^6$ F |
| | 4 | Core Temperature | 1 | 4765.2 J/K |
| | 5 | Surface Temperature | 1 | 342 J/K |
| | 6 | Current | 2 | 0 |
| | $1^s$ | Temperature | - | - |
| | $2^s$ | Temperature | - | - |
| High Voltage Bus | 1 | Bus Voltage | 2 | 1 F |
| | 2 | Current | 2 | 0 |
| | 3 | Current | 2 | 0 |
| | 4 | Current | 2 | 0 |
| | $1^s$ | Current | - | - |
| | $2^s$ | Voltage | - | - |
| | $3^s$ | Voltage | - | - |
| | $4^s$ | Voltage | - | - |
| | $5^s$ | Temperature | - | - |
| Low Voltage Bus | 1 | Bus Voltage | 2 | 1 F |
| | 2 | Current | 2 | 0 |
| | $1^s$ | Voltage | - | - |
| | $2^s$ | Current | - | - |
| | $3^s$ | Current | - | - |
| | $4^s$ | Current | - | - |
| | $5^s$ | Current | - | - |
| | $6^s$ | Current | - | - |
| | $7^s$ | Temperature | - | - |
| | $8^s$ | Current | - | - |
| Inverter (Inv), Rectifier (Rec), Converter (Conv) | 1 | Voltage | 2 | 0.05 F |
| | 2 | Current | 2 | 0 |
| | 3 | Voltage | 2 | 0.05 F |
| | 4 | Temperature | 1 | Inv: 8849 J/K Rec: 6055 J/K Conv: 2950 J/K |
| | $1^s$ | Current | - | - |
| | $2^s$ | Current | - | - |
| | $3^s$ | Temperature | - | - |

Table 2. Electrical component edges.

| | Edge Number | Tail Vertex | Head Vertex | Edge Type | Parameter Value $\alpha_j$ |
|---|---|---|---|---|---|
| Battery | 1 | 1 | 6 | 2 | $76*V_{oc}$ |
| | 2 | 6 | 2 | 3 | 76 |
| | 3 | 6 | 3 | 3 | 76 |
| | 4 | 6 | 4 | 4 | 0.754 Ω |
| | 5 | 2 | 4 | 4 | $2.02*10^4$ Ω$^{-1}$ |
| | 6 | 3 | 4 | 4 | $1.10*10^4$ Ω$^{-1}$ |
| | 7 | 4 | 5 | 5 | 39.2 W/K |
| | 8 | 6 | $1^s$ | 3 | 1 |
| | 9 | 5 | $2^s$ | 5 | 1 W/K |
| High Voltage Bus | 1 | $1^s$ | 1 | 3 | 1 |
| | 2 | 2 | 1 | 3 | 1 |
| | 3 | 1 | 3 | 3 | 1 |
| | 4 | 1 | 4 | 3 | 1 |
| | 5 | $2^s$ | 2 | 3 | 1 |
| | 6 | 3 | $3^s$ | 3 | 1 |
| | 7 | 4 | $4^s$ | 3 | 1 |
| | 8 | 1 | $5^s$ | 4 | $1.0*10^{-4}$ Ω$^{-1}$ |
| | 9 | 2 | $5^s$ | 4 | $1.0*10^{-4}$ Ω |
| | 10 | 3 | $5^s$ | 4 | $1.0*10^{-4}$ Ω |
| | 11 | 4 | $5^s$ | 4 | $1.0*10^{-4}$ Ω |
| Low Voltage Bus | 1 | $1^s$ | 2 | 3 | 1 |
| | 2 | 2 | 1 | 3 | 1 |
| | 3 | 1 | $2^s$ | 3 | 1 |
| | 4 | 1 | $3^s$ | 3 | 1 |
| | 5 | 1 | $4^s$ | 3 | 1 |
| | 6 | 1 | $5^s$ | 3 | 1 |
| | 7 | 1 | $6^s$ | 3 | 1 |
| | 8 | 1 | $7^s$ | 4 | $1.0*10^{-4}$ Ω$^{-1}$ |
| | 9 | 2 | $7^s$ | 4 | $1.0*10^{-4}$ Ω |
| | 10 | 1 | $8^s$ | 3 | 1 |
| Inverter (Inv), Rectifier (Rec), Converter (Conv) | 1 | $1^s$ | 1 | 3 | 1 |
| | 2 | 1 | 2 | 8 | 1 |
| | 3 | 1 | 4 | 4 | $1.0*10^{-4}$ Ω$^{-1}$ |
| | 4 | 2 | 3 | 3 | 1 |
| | 5 | 2 | 4 | 4 | 0.0050 Ω |
| | 6 | 3 | $2^s$ | 3 | 1 |
| | 7 | 3 | 4 | 4 | $1.0*10^{-4}$ Ω$^{-1}$ |
| | 8 | 4 | $3^s$ | 5 | 100 W/K |

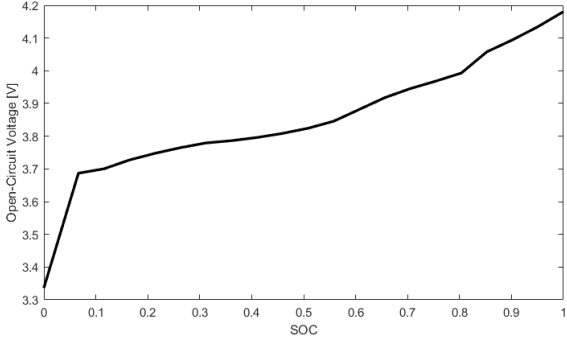

**Figure 2. Open-circuit voltage curve for the battery.**

The second parameter, $f_{w,2}$, captures rolling resistance:

$$f_{w,2} = 0.0178 \cos\left(\theta_g(x_2)\right) f_{sig}(x^{tail}) \quad (3)$$

The variable $\theta_g$ is the gradient of the road, and is a function of distance traveled $x_2$. The third parameter, $f_{w,3}$, models the effect of the road grade on the dynamics:

$$f_{w,3} = 3.14 \sin\left(\theta_g(x_2)\right) \quad (4)$$

The fourth parameter, $f_{w,4}$, represents the brake force acting on the vehicle:

$$f_{w,4} = 3840 f_{sig}(x_{tail}) \quad (5)$$

The fifth parameter, $f_{w,5}$, is used to calculate the distance the vehicle travels over time, with $x_1$ as the wheel's angular velocity:

$$f_{w,5} = 0.32 x_1 \quad (6)$$

### C. Thermal Management System Components

Components with primarily or entirely thermal dynamics include the air cooling path passing through the vehicle, the cabin and its air path, and the vapor compression system (VCS). This also includes liquid cooling loops for the power electronics, battery, VCS, and transaxle components. Table 5 and Table 6 present descriptions of the vertices and edges, respectively. Figure 5 presents the thermal graphs.

The cabin air volume is drawn from [2], and is used determine equivalent edge and capacitance parameter values. The mass flow rates are functions of the corresponding fan or pump angular velocity $\omega$, using scaled mass flow rates from [4] as a baseline. The maps for the mass flow rates of air, $\dot{m}_a$, and the liquid, $\dot{m}_l$, are given by Eq. (7) and Eq. (8), respectively. The mass flow rates act as inputs to edges, with units of kg/s.

$$\dot{m}_a = 0.0139 + 0.0137 \omega \quad (7)$$

$$\dot{m}_l = 0.0040 \omega \quad (8)$$

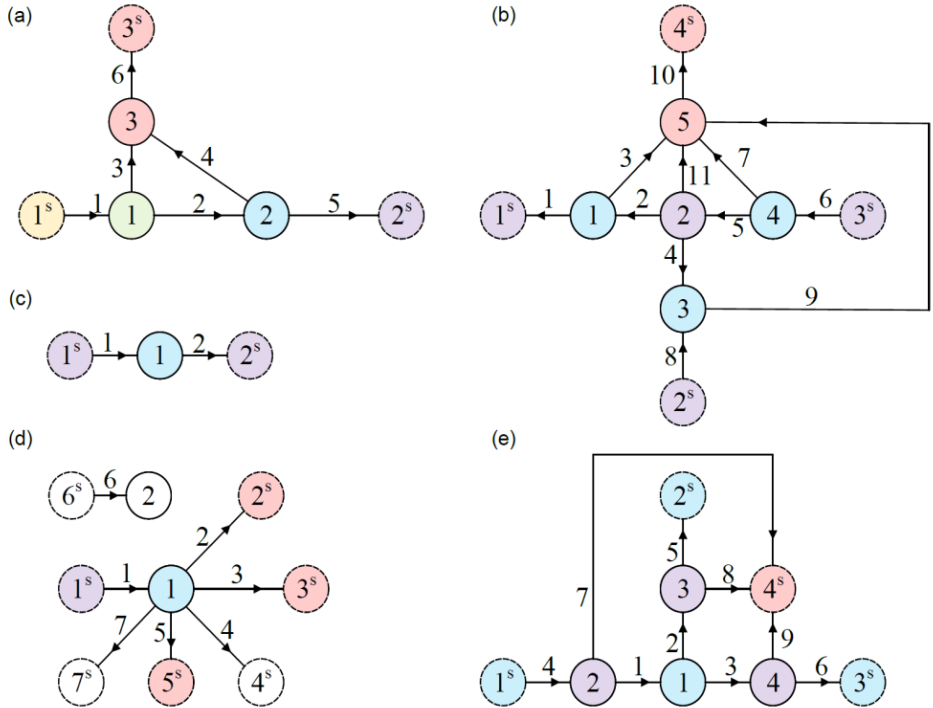

**Figure 3: Graph-based models for (a) a motor, generator, pump or fan, (b) a planetary gear, (c) an engine, (d) wheels and vehicle chassis, and (e) a transmission.**

Table 3. Mechanical component vertices.

| | Vertex Number | State | Vertex Type | Parameter Value $c_i$ |
|---|---|---|---|---|
| Motor (Mot), Pump, Fan, Generator (Gen) | 1 | Current | 2 | 0.05 H |
| | 2 | Angular Velocity | 2 | 0.05 kg-m$^2$ |
| | 3 | Temperature | 1 | Mot: 3*10$^4$ J/K Gen: 1.33*10$^4$ J/K Pump: 10 J/K Fan: 10 J/K |
| | 1$^s$ | Voltage | - | - |
| | 2$^s$ | Torque | - | - |
| | 3$^s$ | Temperature | - | - |
| Engine | 1 | Angular Velocity | 2 | 0.18 kg-m$^2$ |
| | 1$^s$ | Torque | - | - |
| | 2$^s$ | Torque | - | - |
| Planetary Gear | 1 | Sun Angular Velocity | 2 | 0 |
| | 2 | Torque | 2 | 0 |
| | 3 | Carrier Angular Velocity | 2 | 0 |
| | 4 | Ring Angular Velocity | 2 | 0 |
| | 5 | Temperature | 1 | 1.33*10$^3$ J/K |
| | 1$^s$ | Torque | - | - |
| | 2$^s$ | Torque | - | - |
| | 3$^s$ | Torque | - | - |
| | 4$^s$ | Temperature | - | - |
| Transmission | 1 | Angular Velocity | 2 | 0 |
| | 2 | Torque | 2 | 0 |
| | 3 | Torque | 2 | 0 |
| | 4 | Torque | 2 | 0 |
| | 1$^s$ | Angular Velocity | - | - |
| | 2$^s$ | Angular Velocity | - | - |
| | 3$^s$ | Angular Velocity | - | - |
| | 4$^s$ | Temperature | - | - |
| Wheels and Chassis | 1 | Wheel Angular Velocity | 2 | 1 kg-m$^2$ |
| | 2 | Distance Traveled | 2 | 1 |
| | 1$^s$ | Torque | - | - |
| | 2$^s$ | Temperature | - | - |
| | 3$^s$ | Temperature | - | - |
| | 4$^s$ | - | - | - |
| | 5$^s$ | Temperature | - | - |
| | 6$^s$ | - | - | - |
| | 7$^s$ | - | - | - |

Table 4. Mechanical component edges.

| | Edge Number | Tail Vertex | Head Vertex | Edge Type | Parameter Value $\alpha_j$ |
|---|---|---|---|---|---|
| Motor (Mot), Pump, Fan, Generator (Gen) | 1 | 1$^s$ | 1 | 3 | 1 |
| | 2 | 1 | 2 | 3 | Mot: 0.15 J/A Gen: 0.15 J/A Pump: 0.25 J/A Fan: 0.25 J/A |
| | 3 | 1 | 3 | 4 | Mot: 0.01 Ω Gen: 0.01 Ω Pump: 0.05 Ω Fan: 0.05 Ω |
| | 4 | 2 | 3 | 4 | Mot: 0 Gen: 0 Pump: 0.05 J-s Fan: 0.05 J-s |
| | 5 | 2 | 2$^s$ | 3 | 1 |
| | 6 | 3 | 3$^s$ | 5 | Mot: 100 W/K Gen: 100 W/K Pump: 1000 W/K Fan: 1000 W/K |
| Engine | 1 | 1$^s$ | 1 | 2 | $\tau_e$ |
| | 2 | 1 | 2$^s$ | 3 | 1 |
| Planetary Gear | 1 | 1 | 1$^s$ | 3 | 1 |
| | 2 | 2 | 1 | 3 | 1 |
| | 3 | 1 | 5 | 4 | 0.01 J-s |
| | 4 | 2 | 3 | 3 | 2.53 |
| | 5 | 4 | 2 | 3 | 3.53 |
| | 6 | 3$^s$ | 4 | 3 | 1 |
| | 7 | 4 | 5 | 4 | 0.01 J-s |
| | 8 | 2$^s$ | 3 | 3 | 1 |
| | 9 | 3 | 5 | 4 | 0.01 J-s |
| | 10 | 5 | 4$^s$ | 5 | 100 W/K |
| | 11 | 2 | 5 | 4 | 0 |
| Transmission | 1 | 2 | 1 | 3 | 3.58 |
| | 2 | 1 | 3 | 3 | 1 |
| | 3 | 1 | 4 | 3 | 1 |
| | 4 | 1$^s$ | 2 | 3 | 1 |
| | 5 | 3 | 2$^s$ | 3 | 0.714 |
| | 6 | 4 | 3$^s$ | 3 | 2.57 |
| | 7 | 2 | 4$^s$ | 4 | 1.0*10$^{-4}$ J$^{-1}$s$^{-1}$ |
| | 8 | 3 | 4$^s$ | 4 | 1.0*10$^{-4}$ J$^{-1}$s$^{-1}$ |
| | 9 | 4 | 4$^s$ | 4 | 1.0*10$^{-4}$ J$^{-1}$s$^{-1}$ |
| Wheels and Chassis | 1 | 1$^s$ | 1 | 3 | 1 kg-m$^2$ |
| | 2 | 1 | 2$^s$ | 6 | $f_{w,1}$ |
| | 3 | 1 | 3$^s$ | 1 | $f_{w,2}$ |
| | 4 | 1 | 4$^s$ | 1 | $f_{w,3}$ |
| | 5 | 1 | 5$^s$ | 7 | $f_{w,4}$ |
| | 6 | 6$^s$ | 3 | 2 | $f_{w,5}$ |
| | 7 | 1 | 7$^s$ | 4 | 0.3468 J-s |

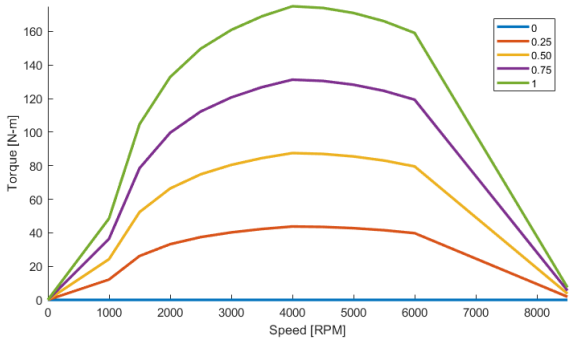

**Figure 4. Engine torque for different throttle commands.**

## D. Connectors

Connector components are used to ensure proper connection between the electrical, mechanical, and thermal component graphs. Virtual inductors are used for electrical-to-electrical connections to maintain proper graph-based modeling format [5]. Gearboxes capture basic dynamics of gears to connect mechanical components. Radiators connect thermal vertices. The thermal combiner is used to connect the thermal vertices of the power electronics. These connection components are presented in Table 7, Table 8, and Figure 6.

## III. THERMO-ELECTRO-MECHANICAL HEV MODEL

This section presents a thermo-electro-mechanical model of an HEV powertrain with cooling. Multiple versions of component graphs are used within the model (e.g., there are multiple DC/DC converters). The numbered component graphs are as follows:

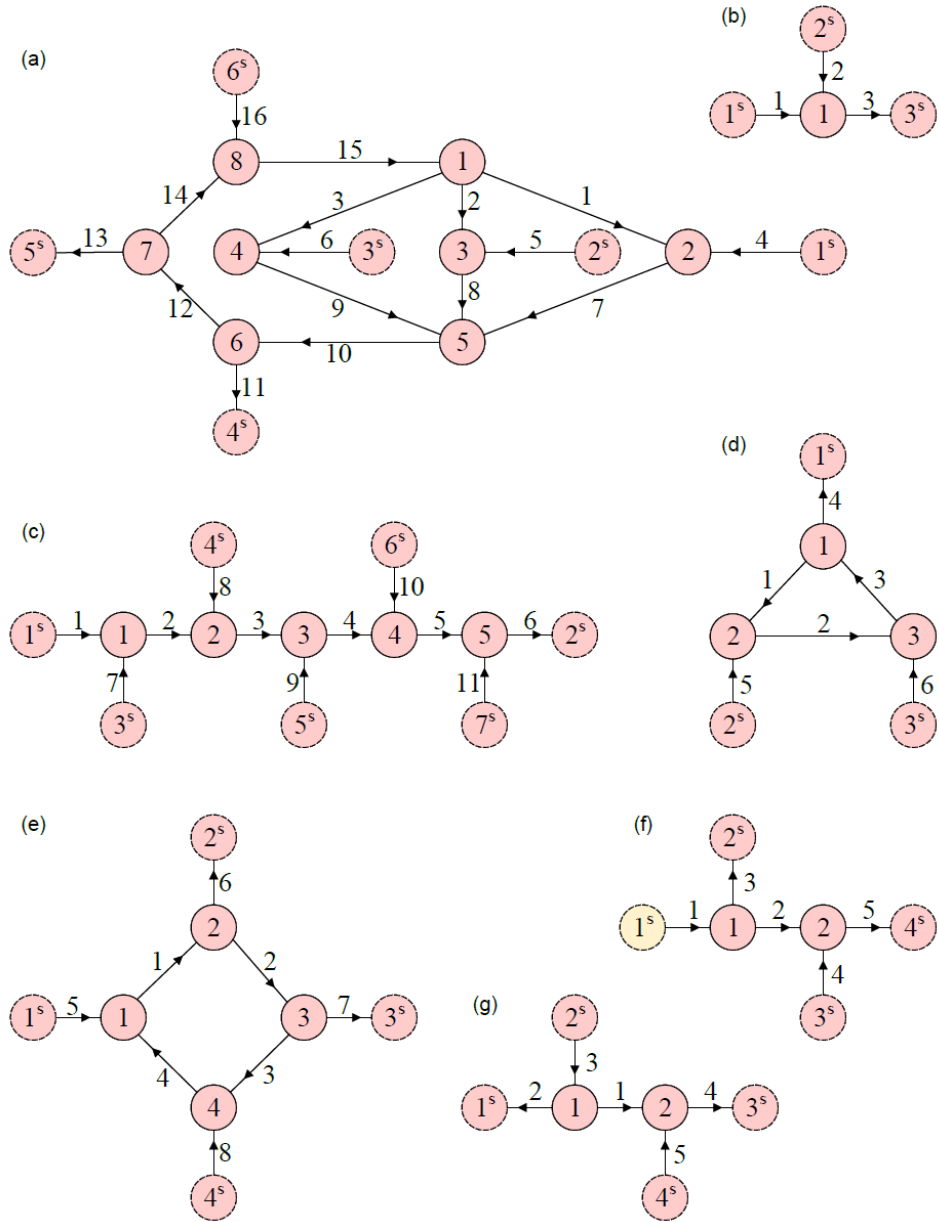

**Figure 5: Graph-based models for (a) a transaxle liquid cooling loop, (b) a cabin, (c) an air path through the vehicle, (d) a battery or VCS liquid cooling loop, (e) a power electronics liquid cooling loop, (f) a VCS, and (g) a cabin inlet air path.**

1. Battery
2. High voltage bus
3. Inverter
4. Motor
5. Transmission
6. Wheels and chassis
7. Converter #1
8. Low voltage bus
9. Air cooling path
10. Power electronics liquid cooling loop
11. Radiator #1
12. Thermal combiner
13. Virtual inductor #1
14. Virtual inductor #2
15. Converter #2
16. Converter #3
17. Pump #1
18. Fan #1
19. VCS
20. Converter #4
21. Virtual inductor #3
22. VCS liquid cooling loop
23. Battery liquid cooling loop
24. Pump #2
25. Converter #5
26. Virtual inductor #4
27. Cabin air path
28. Cabin
29. Fan #2
30. Converter #6
31. Virtual inductor #5
32. Transaxle liquid cooling loop
33. Radiator #2
34. Pump #3
35. Converter #7
36. Virtual inductor #6
37. Rectifier
38. Generator
39. Gearbox #1
40. Gearbox #2
41. Engine
42. Planetary gear

**Table 5. Thermal component vertices.**

| | Vertex Number | State | Vertex Type | Parameter Value $c_i$ |
|---|---|---|---|---|
| Air Cooling Path | 1 | Temperature | 1 | 77.8 J/K |
| | 2 | Temperature | 1 | 77.8 J/K |
| | 3 | Temperature | 1 | 77.8 J/K |
| | 4 | Temperature | 1 | 77.8 J/K |
| | 5 | Temperature | 1 | 77.8 J/K |
| | $1^s$ | Temperature | - | - |
| | $2^s$ | Temperature | - | - |
| | $3^s$ | Temperature | - | - |
| | $4^s$ | Temperature | - | - |
| | $5^s$ | Temperature | - | - |
| | $6^s$ | Temperature | - | - |
| | $7^s$ | Temperature | - | - |
| Power Electronics Liquid Cooling Loop | 1 | Temperature | 1 | $1.34*10^3$ J/K |
| | 2 | Temperature | 1 | 336 J/K |
| | 3 | Temperature | 1 | 671 J/K |
| | 4 | Temperature | 1 | 336 J/K |
| | $1^s$ | Temperature | - | - |
| | $2^s$ | Temperature | - | - |
| | $3^s$ | Temperature | - | - |
| | $4^s$ | Temperature | - | - |
| Cabin | 1 | Temperature | 1 | $3.84*10^3$ J/K |
| | $1^s$ | Temperature | - | - |
| | $2^s$ | Temperature | - | - |
| | $3^s$ | Temperature | - | - |
| Cabin Air Path | 1 | Temperature | 1 | 25.7 J/K |
| | 2 | Temperature | 1 | 25.7 J/K |
| | $1^s$ | Temperature | - | - |
| | $2^s$ | Temperature | - | - |
| | $3^s$ | Temperature | - | - |
| | $4^s$ | Temperature | - | - |
| Vapor Compression System (VCS) | 1 | VCS State 1 | 1 | 1 |
| | 2 | VCS State 2 | 1 | 0 |
| | $1^s$ | Voltage | - | - |
| | $2^s$ | Temperature | - | - |
| | $3^s$ | Temperature | - | - |
| | $4^s$ | Temperature | - | - |
| Battery Liquid Cooling Loop | 1 | Temperature | 1 | 555 J/K |
| | 2 | Temperature | 1 | $1.11*10^3$ J/K |
| | 3 | Temperature | 1 | 278 J/K |
| | $1^s$ | Temperature | - | - |
| | $2^s$ | Temperature | - | - |
| | $3^s$ | Temperature | - | - |
| VCS Liquid Cooling Loop | 1 | Temperature | 1 | $6*10^3$ J/K |
| | 2 | Temperature | 1 | $6*10^3$ J/K |
| | 3 | Temperature | 1 | $6*10^3$ J/K |
| | $1^s$ | Temperature | - | - |
| | $2^s$ | Temperature | - | - |
| | $3^s$ | Temperature | - | - |
| Transaxle Liquid Cooling Loop | 1 | Temperature | 1 | 645 J/K |
| | 2 | Temperature | 1 | $1.29*10^3$ J/K |
| | 3 | Temperature | 1 | $2.58*10^3$ J/K |
| | 4 | Temperature | 1 | $3.87*10^3$ J/K |
| | 5 | Temperature | 1 | 645 J/K |
| | 6 | Temperature | 1 | 645 J/K |
| | 7 | Temperature | 1 | $3.87*10^3$ J/K |
| | 8 | Temperature | 1 | 645 J/K |
| | $1^s$ | Temperature | - | - |
| | $2^s$ | Temperature | - | - |
| | $3^s$ | Temperature | - | - |
| | $4^s$ | Temperature | - | - |
| | $5^s$ | Temperature | - | - |
| | $6^s$ | Temperature | - | - |

**Table 6. Thermal component edges.**

| | Edge Number | Tail Vertex | Head Vertex | Edge Type | Parameter Value $\alpha_j$ |
|---|---|---|---|---|---|
| Air Cooling Path | 1 | $1^s$ | 1 | 1 | 1008 J/(kg-K) |
| | 2 | 1 | 2 | 1 | 1008 J/(kg-K) |
| | 3 | 2 | 3 | 1 | 1008 J/(kg-K) |
| | 4 | 3 | 4 | 1 | 1008 J/(kg-K) |
| | 5 | 4 | 5 | 1 | 1008 J/(kg-K) |
| | 6 | 5 | $2^s$ | 1 | 1008 J/(kg-K) |
| | 7 | $3^s$ | 1 | 5 | 0 |
| | 8 | $4^s$ | 2 | 5 | 0 |
| | 9 | $5^s$ | 3 | 5 | 0 |
| | 10 | $6^s$ | 4 | 5 | 0 |
| | 11 | $7^s$ | 5 | 5 | 0 |
| Power Electronics Liquid Cooling Loop | 1 | 1 | 2 | 1 | 4000 J/(kg-K) |
| | 2 | 2 | 3 | 1 | 4000 J/(kg-K) |
| | 3 | 3 | 4 | 1 | 4000 J/(kg-K) |
| | 4 | 4 | 1 | 1 | 4000 J/(kg-K) |
| | 5 | $1^s$ | 1 | 5 | 905 W/K |
| | 6 | 2 | $2^s$ | 5 | 0 W/K |
| | 7 | 3 | $3^s$ | 5 | 151 W/K |
| | 8 | $4^s$ | 4 | 5 | 100 W/K |
| Cabin | 1 | $1^s$ | 1 | 1 | 1008 J/(kg-K) |
| | 2 | $2^s$ | 1 | 1 | 1 |
| | 3 | 1 | $3^s$ | 1 | 1008 J/(kg-K) |
| Cabin Air Path | 1 | 1 | 2 | 1 | 1008 J/(kg-K) |
| | 2 | 1 | $1^s$ | 5 | 33.3 W/K |
| | 3 | $2^s$ | 1 | 1 | 1008 J/(kg-K) |
| | 4 | 2 | $3^s$ | 1 | 1008 J/(kg-K) |
| | 5 | $4^s$ | 2 | 5 | 1000 W/K |
| Vapor Compression System (VCS) | 1 | $1^s$ | 1 | 4 | 1.64 $\Omega^{-1}$ |
| | 2 | 1 | 2 | 1 | 0.85 |
| | 3 | 1 | $2^s$ | 1 | 0.15 |
| | 4 | $3^s$ | 2 | 2 | 4.16 |
| | 5 | 2 | $4^s$ | 1 | 5.16 |
| Battery Liquid Cooling Loop | 1 | 1 | 2 | 7 | 4000 J/(kg-K) |
| | 2 | 2 | 3 | 7 | 4000 J/(kg-K) |
| | 3 | 3 | 1 | 7 | 4000 J/(kg-K) |
| | 4 | 1 | $1^s$ | 5 | 1000 W/K |
| | 5 | $2^s$ | 2 | 5 | 43.75 W/K |
| | 6 | $3^s$ | 3 | 5 | 1000 W/K |
| VCS Liquid Cooling Loop | 1 | 1 | 2 | 1 | 400 W/K |
| | 2 | 2 | 3 | 1 | 400 W/K |
| | 3 | 3 | 1 | 1 | 400 W/K |
| | 4 | 1 | $1^s$ | 5 | 1000 W/K |
| | 5 | $2^s$ | 2 | 5 | 1000 W/K |
| | 6 | $3^s$ | 3 | 5 | 1000 W/K |
| Transaxle Liquid Cooling Loop | 1 | 1 | 2 | 7 | 1333 J/(kg-K) |
| | 2 | 1 | 3 | 7 | 1333 J/(kg-K) |
| | 3 | 1 | 4 | 7 | 1333 J/(kg-K) |
| | 4 | $1^s$ | 2 | 5 | 1600 W/K |
| | 5 | $2^s$ | 3 | 5 | 53.33 W/K |
| | 6 | $3^s$ | 4 | 5 | 216.7 W/K |
| | 7 | 2 | 5 | 7 | 1333 J/(kg-K) |
| | 8 | 3 | 5 | 7 | 1333 J/(kg-K) |
| | 9 | 4 | 5 | 7 | 1333 J/(kg-K) |
| | 10 | 5 | 6 | 7 | 4000 J/(kg-K) |
| | 11 | 6 | $4^s$ | 5 | 0 W/K |
| | 12 | 6 | 7 | 7 | 4000 J/(kg-K) |
| | 13 | 7 | $5^s$ | 5 | 1 W/K |
| | 14 | 7 | 8 | 7 | 4000 J/(kg-K) |
| | 15 | 8 | 1 | 7 | 4000 J/(kg-K) |
| | 16 | $6^s$ | 8 | 5 | 1 W/K |

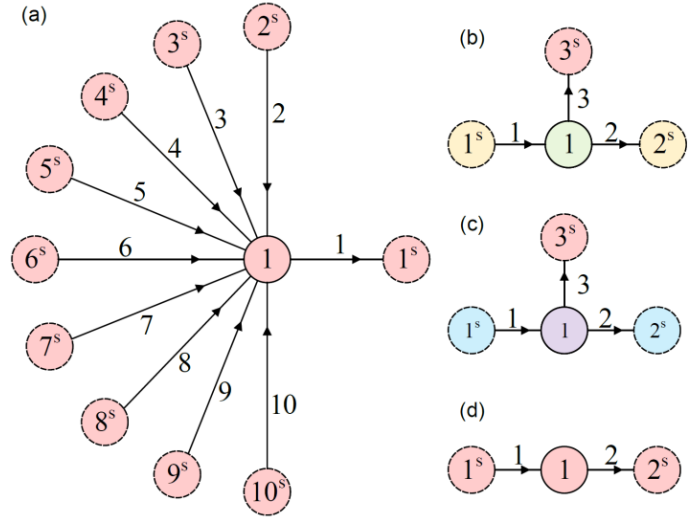

Figure 6. Graph-based models for (a) a thermal combiner, (b) a virtual inductor, (c) a gearbox, and (d) a radiator.

Table 7. Connector component vertices.

| | Vertex Number | State | Vertex Type | Parameter Value $c_i$ |
|---|---|---|---|---|
| Virtual Inductor | 1 | Current | 2 | 0 |
| | $1^s$ | - | - | - |
| | $2^s$ | - | - | - |
| | $3^s$ | - | - | - |
| Gearbox | 1 | Torque | 2 | 0 |
| | $1^s$ | - | - | - |
| | $2^s$ | - | - | - |
| | $3^s$ | - | - | - |
| Radiator | 1 | Temperature | 1 | 1 J/K |
| | $1^s$ | - | - | - |
| | $2^s$ | - | - | - |
| Thermal Combiner | 1 | Temperature | 1 | 884.9 J/K |
| | $1^s$ | - | - | - |
| | $2^s$ | - | - | - |
| | $3^s$ | - | - | - |
| | $4^s$ | - | - | - |
| | $5^s$ | - | - | - |
| | $6^s$ | - | - | - |
| | $7^s$ | - | - | - |
| | $8^s$ | - | - | - |
| | $9^s$ | - | - | - |
| | $10^s$ | - | - | - |

Table 8. Connector component edges.

| | Edge Number | Tail Vertex | Head Vertex | Edge Type | Parameter Value $\alpha_j$ |
|---|---|---|---|---|---|
| Virtual Inductor | 1 | $1^s$ | 1 | 3 | 1 |
| | 2 | 1 | $2^s$ | 3 | 1 |
| | 3 | 1 | $3^s$ | 4 | $1.0*10^{-4}$ Ω |
| Gearbox | 1 | $1^s$ | 1 | 3 | 1 |
| | 2 | 1 | $2^s$ | 3 | 1 |
| | 3 | 1 | $3^s$ | 4 | 0 |
| Radiator | 1 | $1^s$ | 1 | 5 | 1000 W/K |
| | 2 | 1 | $2^s$ | 5 | 1000 W/K |
| Thermal Combiner | 1 | 1 | $1^s$ | 5 | 100 W/K |
| | 2 | $2^s$ | 1 | 5 | 100 W/K |
| | 3 | $3^s$ | 1 | 5 | 100 W/K |
| | 4 | $4^s$ | 1 | 5 | 100 W/K |
| | 5 | $5^s$ | 1 | 5 | 100 W/K |
| | 6 | $6^s$ | 1 | 5 | 100 W/K |
| | 7 | $7^s$ | 1 | 5 | 100 W/K |
| | 8 | $8^s$ | 1 | 5 | 100 W/K |
| | 9 | $9^s$ | 1 | 5 | 100 W/K |
| | 10 | $10^s$ | 1 | 5 | 100 W/K |

The component graphs are combined into the full HEV model by defining which edges from which component graphs are merged into one [6]. The respective head and tail vertices of merged edges are also merged. Table 9 defines the edges that are merged, as well as the dominant edge, which is the effective edge type and parameters used in the full model. Figures 7 and 8 present the configuration of the full thermo-electro-mechanical model of the power-split HEV powertrain with cooling.

## IV. PLANT DESGIN VARIABLES

This section presents the physical design variables to be adjusted and optimized for the HEV model. Each design variable impacts capacitance and edge parameters through the $\mathbf{\Psi_c}$ and $\mathbf{\Psi}$ matrices of an augmented graph-based model [7]. Six plant design variables are defined:

1. $\theta_{c,1}$ is size of the heat exchanger near the battery. To determine the equivalent terms for $\mathbf{\Psi_c}$ and $\mathbf{\Psi}$, component graph 23's 2nd capacitance is multiplied by $\theta_{c,1}$, as well as edge 5's parameter.
2. $\theta_{c,2}$ is size of the heat exchanger near the power electronics. To determine the equivalent terms for $\mathbf{\Psi_c}$ and $\mathbf{\Psi}$, component graph 10's 1st capacitance is multiplied by $\theta_{c,2}$, as well as edge 5's parameter.
3. $\theta_{c,3}$ is size of the heat exchanger near the motor. To determine the equivalent terms for $\mathbf{\Psi_c}$ and $\mathbf{\Psi}$, component graph 32's 4th capacitance is multiplied by $\theta_{c,3}$, as well as edge 6's parameter.

**Table 9. Graphical connections for full thermo-electro-mechanical HEV model. Shaded boxes represent the graph with the dominant edge.**

| Connection Number | 1st Component Graph | 1st Component Edge | 2nd Component Graph | 2nd Component Edge |
|---|---|---|---|---|
| 1 | 1 | 8 | 2 | 1 |
| 2 | 2 | 6 | 3 | 1 |
| 3 | 3 | 6 | 4 | 1 |
| 4 | 4 | 5 | 5 | 4 |
| 5 | 5 | 6 | 6 | 1 |
| 6 | 2 | 7 | 7 | 1 |
| 7 | 7 | 6 | 8 | 1 |
| 8 | 9 | 7 | 11 | 2 |
| 9 | 11 | 1 | 10 | 7 |
| 10 | 18 | 6 | 9 | 11 |
| 11 | 17 | 6 | 10 | 8 |
| 12 | 24 | 6 | 23 | 6 |
| 13 | 34 | 6 | 32 | 16 |
| 14 | 17 | 1 | 15 | 6 |
| 15 | 18 | 1 | 16 | 6 |
| 16 | 19 | 1 | 20 | 6 |
| 17 | 24 | 1 | 25 | 6 |
| 18 | 29 | 1 | 30 | 6 |
| 19 | 34 | 1 | 35 | 6 |
| 20 | 8 | 3 | 13 | 1 |
| 21 | 8 | 4 | 14 | 1 |
| 22 | 8 | 5 | 21 | 1 |
| 23 | 8 | 6 | 26 | 1 |
| 24 | 8 | 7 | 31 | 1 |
| 25 | 8 | 10 | 36 | 1 |
| 26 | 15 | 1 | 13 | 2 |
| 27 | 16 | 1 | 14 | 2 |
| 28 | 20 | 1 | 21 | 2 |
| 29 | 25 | 1 | 26 | 2 |
| 30 | 30 | 1 | 31 | 2 |
| 31 | 35 | 1 | 36 | 2 |
| 32 | 12 | 1 | 10 | 5 |
| 33 | 12 | 3 | 3 | 8 |
| 34 | 12 | 4 | 7 | 8 |
| 35 | 12 | 5 | 15 | 8 |
| 36 | 12 | 6 | 16 | 8 |
| 37 | 12 | 7 | 20 | 8 |
| 38 | 12 | 8 | 25 | 8 |
| 39 | 12 | 9 | 30 | 8 |
| 40 | 12 | 10 | 35 | 8 |
| 41 | 12 | 2 | 37 | 8 |
| 42 | 9 | 10 | 19 | 5 |
| 43 | 19 | 4 | 22 | 4 |
| 44 | 22 | 5 | 23 | 4 |
| 45 | 22 | 6 | 27 | 2 |
| 46 | 27 | 4 | 28 | 1 |
| 47 | 29 | 6 | 27 | 5 |
| 48 | 23 | 5 | 1 | 9 |
| 49 | 32 | 6 | 4 | 6 |
| 50 | 32 | 5 | 38 | 6 |
| 51 | 32 | 4 | 42 | 10 |
| 52 | 32 | 13 | 33 | 1 |
| 53 | 9 | 8 | 33 | 2 |
| 54 | 2 | 5 | 37 | 1 |
| 55 | 37 | 6 | 38 | 1 |
| 56 | 38 | 5 | 39 | 2 |
| 57 | 39 | 1 | 42 | 1 |
| 58 | 41 | 2 | 40 | 1 |
| 59 | 40 | 2 | 42 | 6 |
| 60 | 42 | 8 | 5 | 5 |

4. $\theta_{c,4}$ is size of the heat exchanger near the generator. To determine the equivalent terms for $\Psi_c$ and $\Psi$, component graph 32's 3rd capacitance is multiplied by $\theta_{c,4}$, as well as edge 5's parameter.

5. $\theta_{c,5}$ is size of the heat exchanger near the planetary gear box. To determine the equivalent terms for $\Psi_c$ and $\Psi$, component graph 32's 2nd capacitance is multiplied by $\theta_{c,5}$, as well as edge 4's parameter.

6. $\theta_{c,6}$ is the number of battery cells in parallel within the battery pack. To determine the equivalent terms for $\Psi_c$ and $\Psi$, the first 5 battery capacitances of component graph 1 are multiplied by $\theta_{c,6}$, as well as the parameters of edges 5-7 and 9. Edge 4's parameter is multiplied by $1/\theta_{c,6}$. Furthermore, the wheels and chassis graph (component graph 6) is also scaled to account for the increase in vehicle mass due to increasing the battery pack size. The 1st capacitance is scaled according to $f_1$ in Eq. (9), and the parameters of edges 3 and 4 are scaled according to $f_2$ in Eq. (10).

$$f_1 = 187 + 0.591\theta_{c,6} \quad (9)$$

$$f_2 = 1808 + 5.776\theta_{c,6} \quad (10)$$

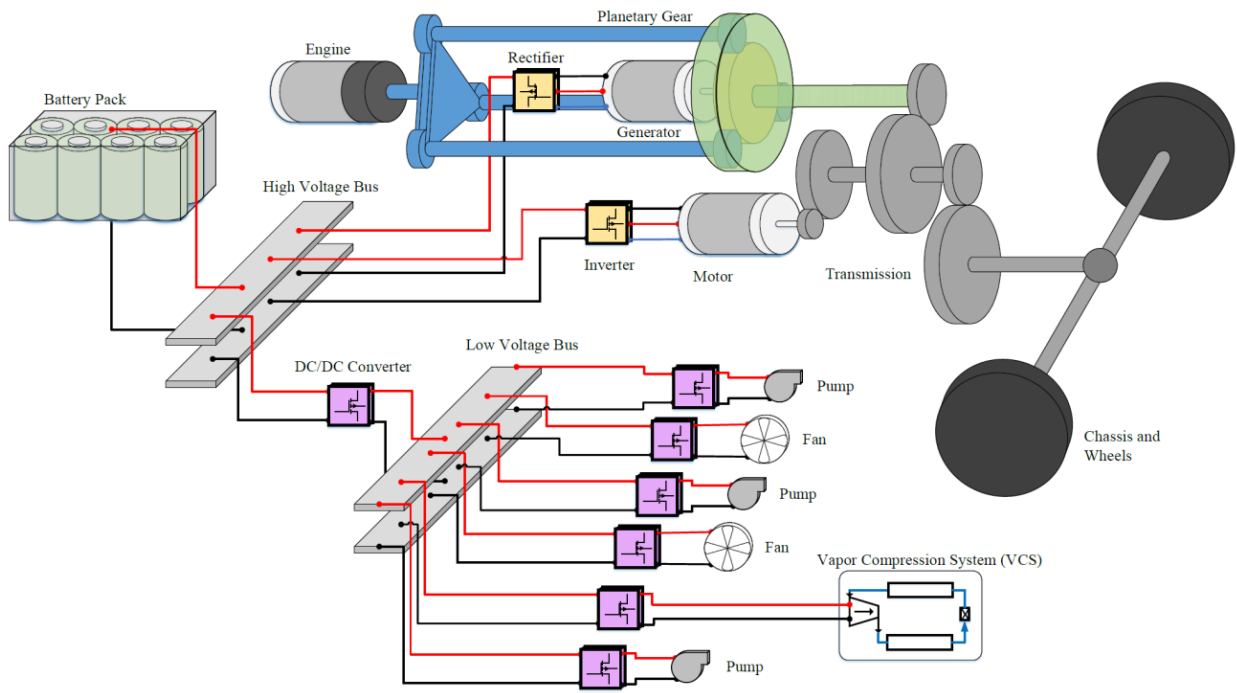

**Figure 7: Electro-mechanical configuration of the HEV powertrain.**

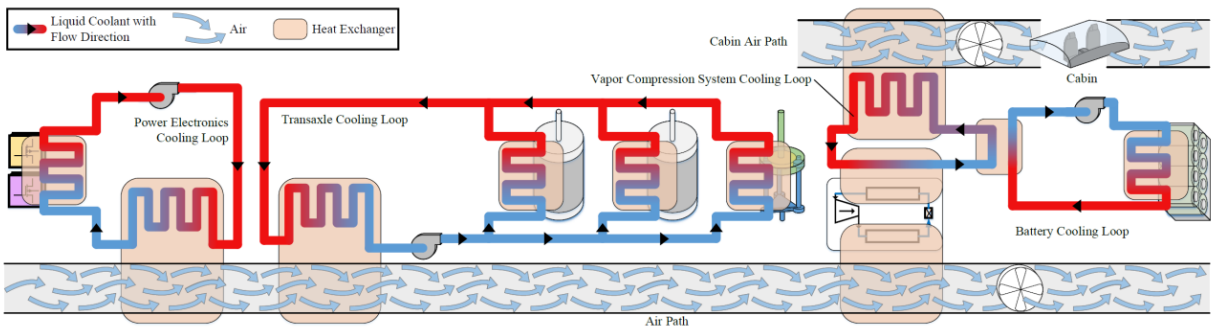

**Figure 8: Thermal cooling system for the HEV.**

23